\documentclass[sigconf]{acmart}

\AtBeginDocument{%
  \providecommand\BibTeX{{%
    \normalfont B\kern-0.5em{\scshape i\kern-0.25em b}\kern-0.8em\TeX}}}

\copyrightyear{2023}
\acmYear{2023}
\setcopyright{acmlicensed}
\acmConference[ISCA '23] {Proceedings of the 50th Annual International Symposium on Computer Architecture}{June 17--21, 2023}{Orlando, FL, USA.}
\acmBooktitle{Proceedings of the 50th Annual International Symposium on Computer Architecture (ISCA '23), June 17--21, 2023, Orlando, FL, USA}
\acmPrice{15.00}
\acmISBN{979-8-4007-0095-8/23/06}
\acmDOI{10.1145/3579371.3589115}

\usepackage{amsmath,amsfonts}
\usepackage{algorithmic}
\usepackage{graphicx}
\usepackage{textcomp}
\usepackage{xcolor}
\usepackage{url}
\usepackage{hyperref}
\usepackage{wrapfig,lipsum,booktabs}
\usepackage{mathtools}
\usepackage{multirow} 
\usepackage{makecell}
\usepackage{epsfig}
\usepackage{fancyhdr}
\usepackage[normalem]{ulem}
\usepackage{soul}
\usepackage{dblfloatfix}

\usepackage{pifont}

\definecolor{my_green}{HTML}{006633}

\newcommand\mynuma[1]{\ifcase#1 \or \ding{172}\or \ding{173}\or
  \ding{174}\or \ding{175}\or \ding{176}\or \ding{177}%
  \or \ding{178}\or \ding{179}\or \ding{180}\or \ding{181}\else *\fi\relax}
\newcommand\mynumb[1]{\ifcase#1 \or \ding{182}\or \ding{183}\or
  \ding{184}\or \ding{185}\or \ding{186}\or \ding{187}%
  \or \ding{188}\or \ding{189}\or \ding{190}\or \ding{191}\else *\fi\relax}
\newcommand{\norm}[1]{\left\lVert#1\right\rVert}

\begin{document}

\title{Instant-3D: Instant Neural Radiance Field Training Towards On-Device AR/VR 3D Reconstruction}
\newcommand{\FrameworkName}{Instant-3D}

\author{Sixu Li}
\authornote{Equal contribution.}
\affiliation{%
  \institution{Georgia Institute of Technology}
  \city{Atlanta, GA}
  \country{USA}
}
\email{sli941@gatech.edu}

\author{Chaojian Li}
\authornotemark[1]
\affiliation{%
  \institution{Georgia Institute of Technology}
  \city{Atlanta, GA}
  \country{USA}
}
\email{cli851@gatech.edu}

\author{Wenbo Zhu}
\affiliation{%
  \institution{Georgia Institute of Technology}
  \city{Atlanta, GA}
  \country{USA}
}
\email{eiclab.gatech@gmail.com}

\author{Boyang (Tony) Yu}
\affiliation{%
  \institution{Georgia Institute of Technology}
  \city{Atlanta, GA}
  \country{USA}
}
\email{eiclab.gatech@gmail.com}

\author{Yang (Katie) Zhao}
\affiliation{%
  \institution{Georgia Institute of Technology}
  \city{Atlanta, GA}
  \country{USA}
}
\email{eiclab.gatech@gmail.com}

\author{Cheng Wan}
\affiliation{%
  \institution{Georgia Institute of Technology}
  \city{Atlanta, GA}
  \country{USA}
}
\email{cwan39@gatech.edu}

\author{Haoran You}
\affiliation{%
  \institution{Georgia Institute of Technology}
  \city{Atlanta, GA}
  \country{USA}
}
\email{hyou37@gatech.edu}

\author{Huihong Shi}
\affiliation{%
  \institution{Georgia Institute of Technology}
  \city{Atlanta, GA}
  \country{USA}
}
\email{eiclab.gatech@gmail.com}

\author{Yingyan (Celine) Lin}
\affiliation{%
  \institution{Georgia Institute of Technology}
  \city{Atlanta, GA}
  \country{USA}
}
\email{celine.lin@gatech.edu}

\renewcommand{\shortauthors}{Li, et al.}

\begin{abstract}
Neural Radiance Field (NeRF) based 3D reconstruction is highly desirable for immersive Augmented and Virtual Reality (AR/VR) applications, but achieving instant (i.e., $<$ 5 seconds) on-device NeRF training remains a challenge. In this work, we first identify the inefficiency bottleneck: the need to interpolate NeRF embeddings up to 200,000 times from a 3D embedding grid during each training iteration. To alleviate this, we propose Instant-3D, an algorithm-hardware co-design acceleration framework that achieves instant on-device NeRF training. Our \ul{algorithm} decomposes the embedding grid representation in terms of color and density, enabling computational redundancy to be squeezed out by adopting different (1) grid sizes and (2) update frequencies for the color and density branches. Our \ul{hardware} accelerator further reduces the dominant memory accesses for embedding grid interpolation by (1) mapping multiple nearby points' memory read requests into one during the feed-forward process, (2) merging embedding grid updates from the same sliding time window during back-propagation, and (3) fusing different computation cores to support the different grid sizes needed by the color and density branches of Instant-3D algorithm. Extensive experiments validate the effectiveness of Instant-3D, achieving a large training time reduction of 41$\times$ - 248$\times$ while maintaining the same reconstruction quality. Excitingly, Instant-3D has enabled instant 3D reconstruction for AR/VR, requiring a reconstruction time of only 1.6 seconds per scene and meeting the AR/VR power consumption constraint of 1.9 W.

\end{abstract}



\maketitle

\section{Introduction}

On-the-fly 3D reconstruction has become a fundamental task in numerous augmented and virtual reality (AR/VR) applications which involve fast-changing environments \cite{dawood200919,zhao2020deja,fassi2016vr}, e.g., virtual room planner \cite{vr_room}, VR painting \cite{vr_paint}, metaverse 3D asset creation \cite{3d_assets}, and virtual telepresence \cite{meta_telepresence}. Specifically, 3D reconstruction takes 2D images from a set of \textit{sparsely sampled views} of a 3D scene as its inputs and then generates images of the same scene from \textit{any desired new view}. 
Compared to offloading to the cloud, on-the-fly 3D reconstruction can offer a smaller communication data size and enhanced privacy protection. For example, a 20 MB reconstructed model~\cite{mildenhall2020nerf} may be used instead of 120 MB jpeg images~\cite{dai2017scannet}. This alternative is critical for application scenarios with unstable or unavailable internet connections, such as virtual telepresence~\cite{meta_telepresence} under which each attendee's environment needs to be reconstructed under varying internet bandwidths at a latency of $<$ 2 seconds \cite{miller1968response,nah2004study}.

Among the tremendously growing efforts devoted to pushing forward the achievable quality of 3D reconstruction, neural radiance field (NeRF)-based reconstruction has stood out~\cite{mildenhall2020nerf} thanks to its state-of-the-art (SOTA) performance in terms of photorealistic reconstruction quality. 
However, while instant (i.e., $<$ 5 seconds \cite{muller2022instant}) NeRF-based reconstruction on AR/VR devices for new scenes is highly desirable in unleashing the big promise of photorealistic 3D reconstruction in many emerging applications, it is still not possible even using the most efficient SOTA NeRF training algorithm~\cite{muller2022instant}.

\begin{figure}[t]
  \centering
  \includegraphics[width=0.98\linewidth]{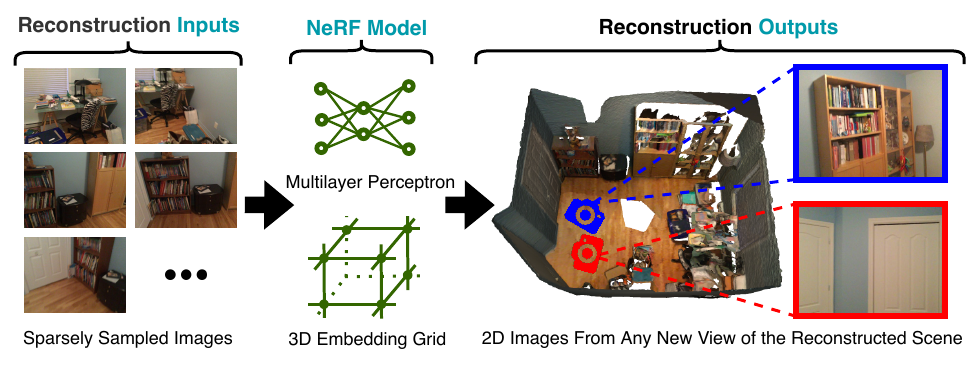}
\vspace{-1.5em}
\caption{An illustration of NeRF-based 3D reconstruction, which takes 2D images from a set of \textit{sparsely sampled views} of a 3D scene as its inputs and then generates images of the same scene from \textit{any desired new view}.
\vspace{-1.2em}
}
\label{fig:3d_reconstruction}
\end{figure}

To close the aforementioned gap, we first set out to understand and identify the key bottleneck that has limited the achievable runtime efficiency of training NeRF-based reconstruction for new scenes. 
To do so, we start by conducting extensive profiling measurements of the most efficient NeRF training algorithm, called Instant-NGP \cite{muller2022instant}, on multiple commercial devices with varying levels of power consumption (e.g., 10 W $\sim$ 20 W). After that, we perform various analyses on the runtime breakdown of each step in Instant-NGP \cite{muller2022instant}'s training pipeline and locate the key bottleneck: the step of interpolating NeRF embeddings from a 3D embedding grid and its corresponding back-propagation process. The purpose of this step is to generate the embeddings of 3D points in the scene to be reconstructed, which include the corresponding points' color and density information, during the reconstruction process. This step needs to be executed more than 200,000 times during each training iteration to ensure photorealistic reconstruction quality.
In particular, the heavy workload of this step dominates around 80\% of the training runtime of the whole pipeline, as discussed and analyzed in Sec.~\ref{sec:SW_ProfilingInstNGP}.

To tackle the inefficiency bottleneck identified above, we develop an algorithm-hardware co-design acceleration framework, and make the following contributions:

\begin{itemize}
    \item We comprehensively profile and analyze the runtime bottleneck in the pipeline of the most efficient NeRF training algorithm \cite{muller2022instant} on multiple devices, and identify the primary cause of the inefficiency: the necessity of interpolating NeRF embeddings from a 3D embedding grid $>$ 200,000 times per training iteration. 

    \item We develop an algorithm-hardware co-design acceleration framework called \textit{Instant-3D} for training NeRFs, which to the best of our knowledge is \textbf{the first} that has achieved \textit{instant} on-device NeRF-based 3D reconstruction. Specifically, Instant-3D targets resolving the aforementioned inefficiency by developing dedicated algorithm and hardware innovations that compress the storage requirement, the number of computations, and the number of accesses for the 3D embedding grid in the identified bottleneck step of embedding grid interpolation. 
    
   \item \ul{On the algorithm level}, leveraging our discovery that color and density have different sensitivities when it comes to compressing NeRFs, we propose to \textbf{decompose the aforementioned embedding grid} in the identified bottleneck in terms of color and density, and then adopt different (1) \textbf{grid sizes} and (2) \textbf{update frequencies} for the resulting color and density branches, allowing for orthogonally squeezing out the computational redundancy in both branches without compromising the reconstruction quality.
    
    \item \ul{On the hardware level}, we propose a dedicated accelerator that leverages the properties of the aforementioned algorithm to boost hardware efficiency while further reducing the dominant memory accesses during the required embedding grid interpolation of NeRFs. 
    The latter leverages our finding that the memory access pattern during embedding grid interpolation is predictable within a specific region of the scene to be reconstructed, as shown in Sec.~\ref{sec:HP_Analysis}. 
    Specifically, our Instant-3D accelerator highlights (1) \textbf{a feed-forward read mapper} that maps the memory read requests of the embeddings of multiple nearby points into one read request during the feed-forward process of NeRF training, (2) \textbf{a back-propagation update merger} that merges multiple embedding grid updates from the same sliding time window into one update during the back-propagation phase of NeRF training, and (3) \textbf{a multi-core-fusion-based reconfigurable scheme} to fuse different computation cores for supporting the different grid sizes needed by the color and density branches of the Instant-3D algorithm.
    
   \item Benchmarking experiments and ablation studies on NeRF-Synthetic \cite{mildenhall2020nerf}, SILVR \cite{courteaux2022silvr}, and ScanNet \cite{dai2017scannet} consistently validate the effectiveness of Instant-3D, achieving a training run reduction of 41$\times$ - 248$\times$ while maintaining the same reconstruction quality as the most efficient NeRF training solution.
   Excitingly, Instant-3D has enavked instant on-device NeRF-based 3D reconstruction for AR/VR, requiring only 1.6 seconds per scene to reach a decent reconstruction PSNR of 25 (acceptable for image representations~\cite{thomos2005optimized,li2007robust}) on NeRF-Synthetic~\cite{mildenhall2020nerf} dataset while meeting the AR/VR power consumption constraint of 1.9 W. 
   
\end{itemize}  
\section{Background and Motivation}
\label{sec:background_motivations}
\subsection{Preliminaries of NeRFs}
\label{sec:nerf_preliminaries}

\begin{figure}[!t]
  \centering
  \includegraphics[width=1.0\linewidth]{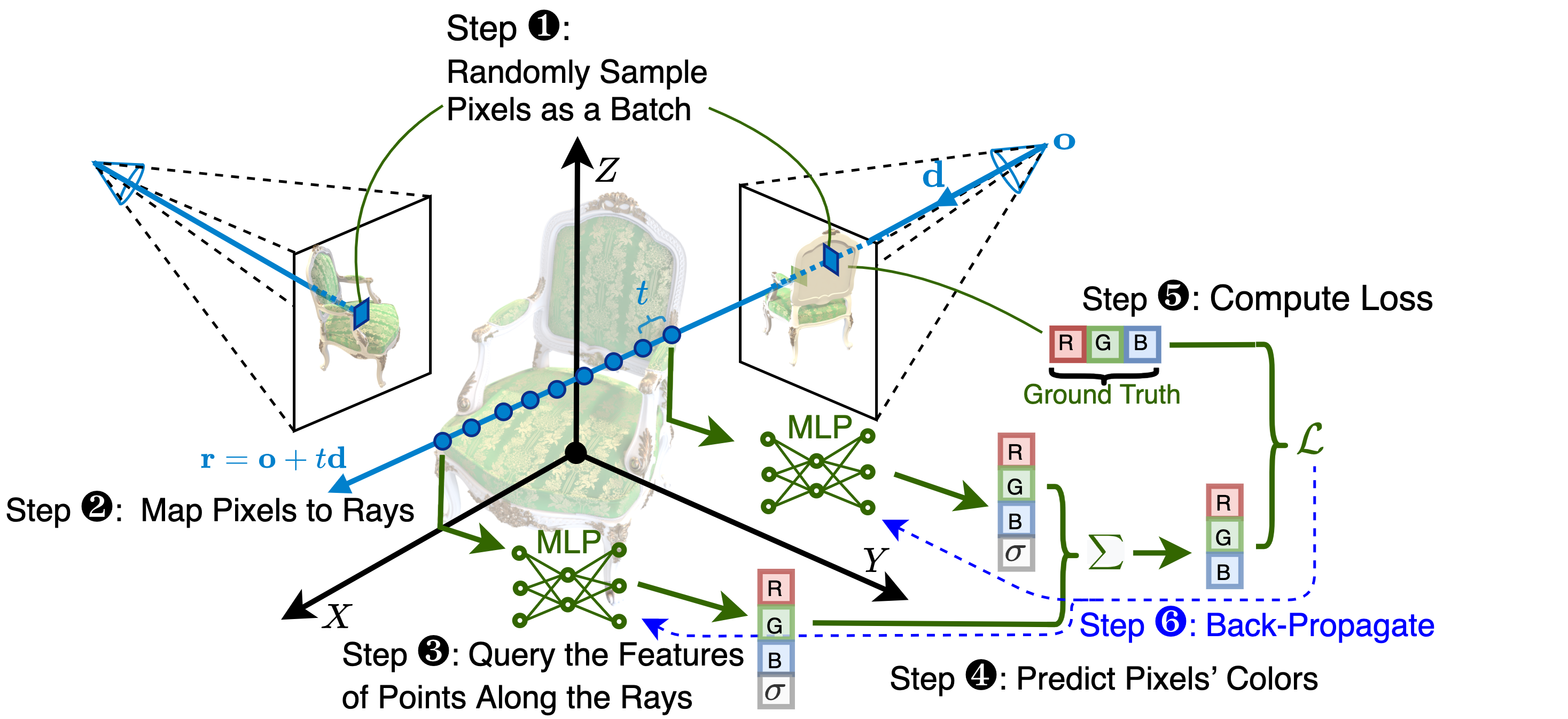}
\vspace{-2em}
\caption{NeRF ~\cite{mildenhall2020nerf}'s training process involves a total of six steps: Step
 {\color{black}{\ding{182}}} randomly samples pixels as a batch, Step
 {\color{black}{\ding{183}}} maps the sampled pixels to rays $\mathbf{r} = \mathbf{o}+t\mathbf{d}$ by emitting rays to pass through the corresponding pixels, Step {\color{black}{\ding{184}}} queries the features (i.e., the RGB color and the density $\sigma$) of points along the rays by providing their locations and directions as the inputs to an MLP model, Step \color{black}{\ding{185}} predicts the pixels' colors following the principle of classical volume rendering~\cite{max1995optical}, Step \color{black}{\ding{186}} computes the loss as the squared error between the predicted colors and ground truth colors, and Step \color{black}{\ding{187}} back-propagates through the above fully differentiable pipeline.}
\label{fig:nerf_training_pipeline}
\vspace{-1.em}
\end{figure}

\textbf{The Training Pipeline of Vanilla NeRFs.}
To reconstruct a specific 3D scene, NeRF takes a set of sparsely sampled views of the same scene as inputs and optimizes an underlying continuous volumetric scene function, i.e., 
 a multilayer perceptron (MLP) model~\cite{mildenhall2020nerf}.
Specifically, NeRFs' training pipeline involves the following six steps as illustrated in Fig.~\ref{fig:nerf_training_pipeline}. \textbf{Step {\color{black}{\ding{182}}}} randomly samples pixels as a batch: One batch of data during each training iteration consists of pixels randomly sampled from all the training images, and these sampled pixels' coordinates and their corresponding RGB colors are the inputs and the ground truth label of the whole NeRF training pipeline, respectively; \textbf{Step {\color{black}{\ding{183}}}} maps the pixels to rays: For each sampled pixel, based on its coordinate, a ray formulated as $\mathbf{r} = \mathbf{o}+t\mathbf{d}$ is emitted from the origin (i.e., the camera's center) $\mathbf{o}$ of its corresponding training view along direction $\mathbf{d}$ to pass through this particular pixel, where $t$ represents the distance between the sampled point along this ray and the origin $\mathbf{o}$; \textbf{Step {\color{black}{\ding{184}}}} queries features of the points along the rays: For each point that has a distance $t_k$ ($k \in [1, N]$, where $N$ represents the total number of the sampled points along each ray) from $\mathbf{o}$, both its location $\mathbf{o}+t_k\mathbf{d}$ and direction $\mathbf{d}$ are applied to an MLP model as inputs. The MLP model then outputs the corresponding density $\sigma_k$ and an RGB color $\mathbf{c}_k$ as the extracted features of this particular point, i.e., $(\mathbf{o}+t_k\mathbf{d}, \mathbf{d}) \rightarrow (\sigma_k, \mathbf{c}_k)$; \textbf{Step {\color{black}{\ding{185}}}} predicts the pixels' colors: Following the principle of classical volume rendering~\cite{max1995optical}, the predicted color $\mathbf{\hat{C}}(\mathbf{r})$ of the pixel corresponding to the ray $\mathbf{r}$ can be computed by integrating the features of all the points along the ray:

\vspace{-1.em}
\begin{align}
\mathbf{\hat{C}}(\mathbf{r}) = \sum_{k=1}^NT_k(1-\exp(-\sigma_k (t_{k+1}-t_{k})))\mathbf{c}_k, \nonumber\\
\textrm{ where } T_k = \exp(-\sum_{j=1}^{k} \sigma_{j} (t_{j+1}-t_{j})),
\label{eq:nerf_render}
\vspace{-0.5em}
\end{align}
where $N$ is the number of sampled points along ray $\mathbf{r}$ and $T_k$ denotes the accumulated transmittance along ray $\mathbf{r}$ to point $\mathbf{o}+t_k\mathbf{d}$, which represents the probability of the ray traveling to the point without hitting any other points; \textbf{Step {\color{black}{\ding{186}}}} computes the loss of reconstructing the scene, which is defined as the total squared error between the predicted colors $\mathbf{\hat{C}}(\mathbf{r})$ and the ground truth colors $\mathbf{C}(\mathbf{r})$:

\vspace{-1.em}
\begin{align}
\mathcal{L}=\sum_{\mathbf{r} \in \mathcal{R}}\left[\norm{\mathbf{\hat{C}}(\mathbf{r})-\mathbf{C}(\mathbf{r})}_2^2\right],
\label{eq:nerf_loss}
\end{align}
where $\mathcal{R}$ is the set of rays in each batch; Finally, \textbf{Step {\color{black}{\ding{187}}}} performs back-propagation: As all of the operations of the previous steps (e.g., Eq.~\ref{eq:nerf_render} and Eq.~\ref{eq:nerf_loss}) are differentiable, the weights of the MLP model in Step {\color{black}{\ding{184}}} can be updated via gradient back-propagation based on the reconstruction loss computed in Step {\color{black}{\ding{186}}}. 

After training, to generate the 2D image corresponding to any desired new view of the reconstructed scene, the only difference from the above pipeline is to replace the randomly sampled pixels in Step {\color{black}{\ding{182}}} with all the pixels in the image of the new view and then stop at Step {\color{black}{\ding{185}}} without further computing the reconstruction loss or back-propagation. 

\textbf{Vanilla NeRFs' Training Cost.} To achieve photorealistic reconstruction quality, the aforementioned training process typically takes around 150,000 iterations per scene, where each iteration executes an MLP model of 1 million FLOPs with a batch size of 786,432 (192 points/pixel $\times$ 4,096 pixels). Therefore, the required total training FLOPs is as large as 353,895 trillion FLOPs, requiring $>$ 1 day of training time on one V100 GPU~\cite{v100}. Such a long training time prohibits vanilla NeRFs~\cite{mildenhall2020nerf} from being used for instant on-device AR/VR scene reconstruction which is highly desirable for many emerging applications, such as photorealistic telepresence~\cite{meta_telepresence}.

\begin{figure}[b]
\vspace{-1.5em}
  \centering
  \includegraphics[width=1.0\linewidth]{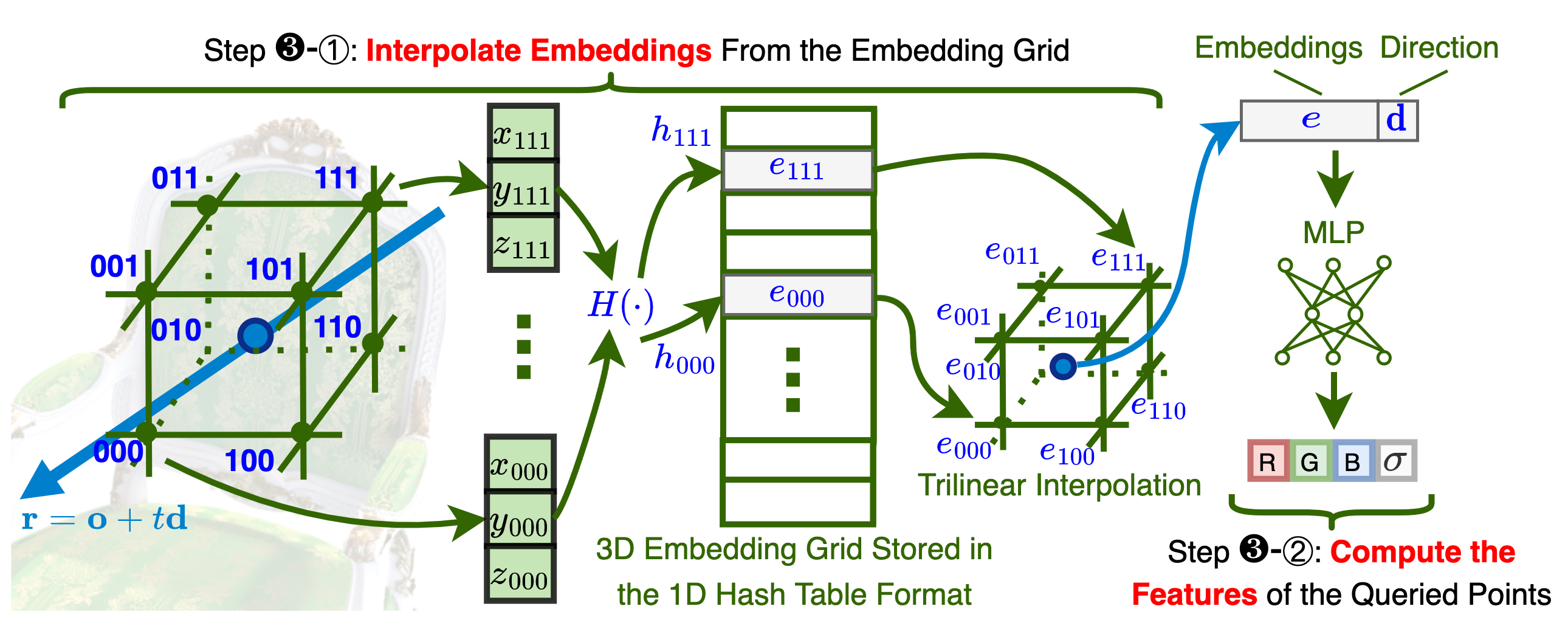}
\vspace{-2em}
\caption{Instant-NGP~\cite{muller2022instant} achieves SOTA training efficiency by replacing Step {\color{black}{\ding{184}}} (i.e., querying the features of points along the rays using a large 10-layer MLP model) in vanilla NeRFs~\cite{mildenhall2020nerf} with both Step {\color{black}{\ding{184}}}-{\color{black}{\ding{172}}} - Interpolating embeddings from the embedding grid and Step {\color{black}{\ding{184}}}-{\color{black}{\ding{173}}} - Computing the features of the queried points using a small MLP model.}
\label{fig:instant_ngp_training_pipeline}
  \vspace{-0.5em}
\end{figure}

\textbf{SOTA Efficient NeRF Training Technique: Instant-NGP.} To alleviate the aforementioned prohibitive cost of training vanilla NeRFs, various works~\cite{muller2022instant,chen2022tensorf,yu2021plenoxels,li2023ingeo} have been proposed to accelerate NeRFs' training process. Among them, Instant-NGP~\cite{muller2022instant} achieved SOTA training speed vs. reconstruction quality trade-offs, e.g., it requires only 5 seconds per scene on a \textit{high-end} RTX3090 GPU~\cite{rtx3090} and thus has been included in all mainstream NeRF-based 3D reconstruction infrastructures~\cite{xrnerf,nerfstudio,li2022nerfacc}. However, Instant-NGP still cannot fulfill the requirement of instant 3D reconstructions on \textit{resource-constrained} AR/VR devices. In particular, to accelerate the NeRF training process, Instant-NGP~\cite{muller2022instant} replaces the MLP model in Step {\color{black}{\ding{184}}}, which is used to query the features of points along the rays, of vanilla NeRFs with a 3D embedding grid; The latter is stored as a compact 1D hash table. In this way, the more costly MLP inference operations (e.g., 1 million FLOPs) in vanilla NeRFs are now converted into much less costly embeddings interpolation operations (e.g., $<$ 0.00005 million FLOPs). Therefore, as visualized in Fig.~\ref{fig:instant_ngp_training_pipeline}, Step {\color{black}{\ding{184}}} in Instant-NGP's training pipeline consists of the following two steps: \textbf{Step {\color{black}{\ding{184}}}-{\color{black}{\ding{172}}}} interpolates embeddings from the embedding grid. Specifically, for each queried point along the rays passing through the pixels of training images, the embeddings $e_{i}$ of its eight nearest vertices $i \in \{000, 001, ..., 111\}$  in the 3D embedding grid will be fetched from the compact 1D hash table of size $T$ through the hash table index $h_i$ based on their coordinates $(x_i, y_i, z_i)$. Finally, the result of the trilinear interpolation on these eight vertices' embeddings will be used as the embeddings of the queried point. In particular, the hash function $H(\cdot)$ that maps the grid vertices' coordinate $(x_i, y_i, z_i)$ to the hash table index $h_i$ is defined as:

\vspace{-1em}
\begin{align}
h_i &= H(x_i, y_i, z_i) = (\pi_1x_i \oplus \pi_2y_i \oplus \pi_3z_i) \textrm{ mod } T,
\label{eq:hash_func}
\end{align}
where $\oplus$ denotes the bit-wise XOR operation, $\pi_1 = 1$, $\pi_2 = 2 654 435 761$, and $\pi_3 = 805 459 861$, following the spatial hash function design in~\cite{teschner2003optimized}. After the queried points, which are along the rays passing through the pixels of training images, obtain their corresponding embeddings from Step {\color{black}{\ding{184}}}-{\color{black}{\ding{172}}}, the embeddings will be applied as the inputs for \textbf{Step {\color{black}{\ding{184}}}-{\color{black}{\ding{173}}}}, which computes the features of the queried points. Specifically, for each queried point, its embeddings $e$ and direction $\mathbf{d}$ are sent as inputs to a small MLP model to obtain the corresponding density $\mathbf{\sigma}$ and view-dependent color $\mathbf{c}$. Here the small MLP only consists of 3 layers with 64 hidden units, in contrast to the required 10 layers with each having 256 hidden units in the vanilla NeRF~\cite{mildenhall2020nerf}. Both the aforementioned lower-cost steps enable Instant-NGP~\cite{muller2022instant}'s SOTA training efficiency. However, even using Instant-NGP~\cite{muller2022instant}, it still requires minutes to days of training time to reconstruct each new scene on edge GPUs~\cite{tx2,xavier_nx,jetson_nano}, which is far from the desired instant runtime (i.e., $<$ 5 seconds~\cite{muller2022instant}) for practical on-device 3D reconstruction.

\begin{figure}[b]
\vspace{-0.5em}
  \centering
  \includegraphics[width=1.0\linewidth]{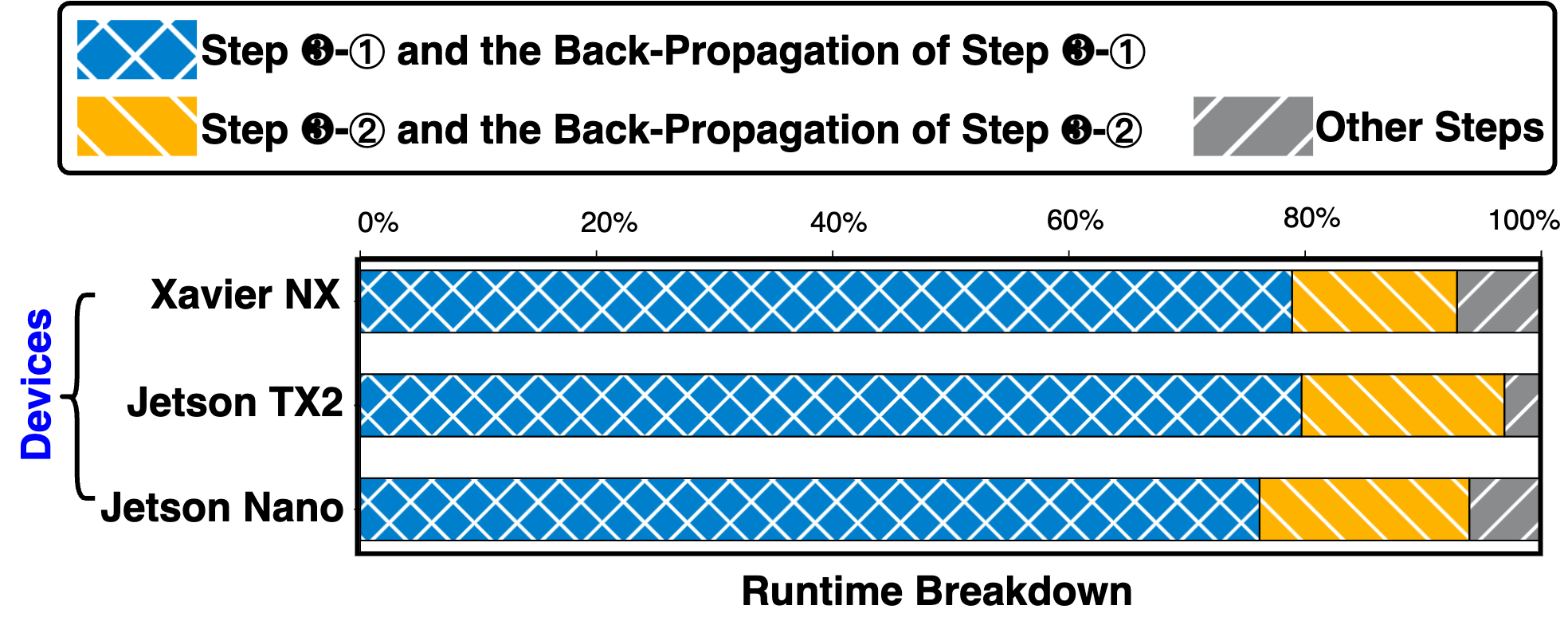}
\vspace{-2.3em}
\caption{Training runtime breakdown averaged on the \textbf{eight scenes} of NeRF-Synthetic~\cite{mildenhall2020nerf} on \textbf{three representative commercial devices}, suggesting that the most efficient NeRF training algorithm~\cite{muller2022instant} is bottlenecked by Step {\color{black}{\ding{184}}}-{\color{black}{\ding{172}}} (i.e., interpolating embeddings from the embedding grid) and its corresponding back-propagation process on all considered scenes and devices.}
\label{fig:profile}
\end{figure}

\subsection{Profiling Analysis of the SOTA Efficient NeRF Training Process}
\label{sec:SW_ProfilingInstNGP}
To close the aforementioned gap between the desired instant on-device 3D reconstruction runtime and the achievable training runtime of the most efficient NeRF training algorithm~\cite{muller2022instant}, we conduct extensive profiling measurements of Instant-NGP~\cite{muller2022instant} on commercial devices with varying levels of power consumption, including Jetson Nano~\cite{jetson_nano}, which typically consumes power consumption of 10 W, Jetson TX2~\cite{tx2}, which typically consumes power consumption of 15 W, and Xavier NX~\cite{xavier_nx}, which typically consumes 20 W. As shown in Fig.~\ref{fig:profile}, the runtime breakdown on the eight commonly used scenes of NeRF-Synthetic~\cite{mildenhall2020nerf} consistently indicates that Step {\color{black}{\ding{184}}}-{\color{black}{\ding{172}}} (i.e., interpolating embeddings from the embedding grid) and its corresponding back-propagation process dominate the overall training runtime of Instant-NGP~\cite{muller2022instant} on all these devices. Based on the above observations, we develop dedicated algorithm and hardware innovations to compress \textit{(1) the storage size of, (2) the number of computations of, and (3) the number of accesses to} the 3D embedding grid in the identified bottleneck of Step {\color{black}{\ding{184}}}-{\color{black}{\ding{172}}} (i.e., interpolating embeddings from the embedding grid). We present our algorithm design in Sec.~\ref{sec:alg} and hardware design in Sec.~\ref{sec:hardware}, respectively.
\section{{\FrameworkName}: Proposed Algorithm}
\label{sec:alg}
In this section, we present our proposed {\FrameworkName} algorithm. First, we hypothesize that the color and density features have different sensitivities when it comes to compressing NeRFs, and thus can evolve at a different pace during NeRF training. 
Our hypothesis has been empirically validated based on the consistent qualitative observations as well as the quantified analyses across different datasets, as discussed in Sec.~\ref{sec:training_speed}. 
Second, to alleviate the training runtime bottleneck during embedding grid interpolation in Instant-NGP, we leverage the above verified sensitivity difference by decoupling the embedding grid into color and density parts.
Specifically, this opens up opportunities to adopt (1) different grid sizes (see Sec.~\ref{sec:different_grid_size}) and (2) different update frequencies (see Sec.~\ref{sec:different_update_frequencies}) for the aforementioned color and density branches of the 3D embedding grid. In this way, the computational redundancy can be squeezed out orthogonally in both branches adaptively to boost the overall training efficiency without compromising the reconstruction quality.

\begin{figure}[b]
  \vspace{-0.3em}
  \centering
  \includegraphics[width=0.9\linewidth]{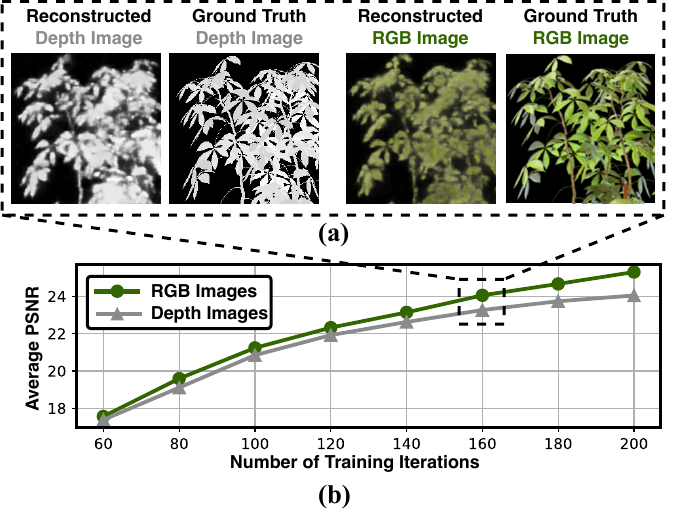}
\vspace{-1.5em}
\caption{(a) \textbf{Color and density feature visualization during training}: Colors are learned faster than the densities under the same number of training iterations. Here we can see that the color features are of higher quality than those of the density under the same number of training iterations (i.e., at the 160th iteration) on the Ficus scene~\cite{mildenhall2020nerf}, where the ground truth color and density features are shown as a reference. (b) \textbf{Quantified PSNR of the color and density features during the whole training trajectory}: The PSNR of the color features is consistently higher than that of the density features during the whole training process. Here the plot shows the average RGB/depth images PSNR on the eight scenes~\cite{mildenhall2020nerf} vs. the number of training iterations.}
\label{fig:color_density_training_speed}
\end{figure}

\subsection{Different Paces of Color and Density During Training}
\label{sec:training_speed}

To validate our hypothesis that the color and density features have different sensitivities to the reconstruction quality of NeRFs, and thus can evolve at a different pace during NeRF training, we have conducted extensive visualization experiments and  analysis. 
Both consistently validate our hypothesis. Here we illustrate and discuss one set of experiment results. Fig.~\ref{fig:color_density_training_speed}(a) visualizes the reconstructed RGB image and depth image of the same scene under the same number of training iterations (here a total of 160). It is worth noting that the depth image is not generated during training and is merely used to test the learned density quality here. We can see that the color information has been better learned than that of the density information under the same training iterations, as the reconstructed RGB color of the ficus is closer to the ground truth one while the fine-grained geometry of the ficus, which is decided by the learned density feature, is still not as clear as compared to its ground truth counterpart.
 Meanwhile, to further quantify the difference regarding the color and density features, Fig.~\ref{fig:color_density_training_speed}(b) shows the Peak Signal-to-Noise Ratio~\cite{hore2010image} (PSNR) (a higher value corresponds to a better-reconstructed quality) of both the reconstructed color and density features during the whole training trajectory. The averaged PSNR vs. the number of training epochs on eight different datasets shows that the color feature is learned at a faster pace than that of density during NeRF training. For example, it takes a total of 160 iterations vs. 200 iterations for the color and density features to evolve to a quality of 24 dB PSNR. 
We conjecture that the aforementioned observations are caused by the different optimization for the color and density, i.e., the training loss (Eq.~\ref{eq:nerf_loss}) is based on the predicted color rather than the predicted density, indicating that optimizing the color is easier and thus the color features are less sensitive to model compression.
Motivated by our above discovery regarding the different sensitivities and pace of the color and density features during NeRF training, we propose to decompose the embedding grid into a color one and a density one. In this way, we can explore the possibility of adopting different grid sizes and different update frequencies for the decomposed branches, as discussed and analyzed in the following two subsections.

\subsection{Different Grid Sizes for the Color and Density Branches}
\label{sec:different_grid_size}

 \begin{wraptable}{r}{0.5\linewidth}
\setlength{\intextsep}{0cm}
\vspace{-1.5em}
\caption{The achieved reconstruction quality (PSNR) vs. the required training time on the eight scenes of NeRF-Synthetic~\cite{mildenhall2020nerf}, when adopting \textbf{different grid sizes} for the density grid $S_D$ and the color grid $S_C$. Here the training runtime is measured on an edge GPU~\cite{xavier_nx}.}
\vspace{-0.5em}
\centering
  \resizebox{1.0\linewidth}{!}
  {
    \begin{tabular}{c||c||c}
    \toprule
    \multirow{2}{*}{$S_D:S_C$} & Average Training & Average \\
     & Runtime (s) & Test PSNR \\
    \midrule
    1:1~\cite{muller2022instant} & 72 & 26.0 \\
    \midrule
    0.25:1  & 65 ($\downarrow$ 9.7\%) & 25.4 \\
    \textbf{1:0.25} & \textbf{63} ($\mathbf{\downarrow}$ \textbf{12.5\%}) & \textbf{26.0} \\
    \bottomrule
    \end{tabular}
    }
  \label{tab:different_grid_sizes}
\vspace{-1em}
\end{wraptable}

\textbf{Observations.} Leveraging the discovery above and our proposed decomposition of the embedding grid, we propose to adopt different grid sizes for the resulting color and density branches of the decomposed embedding grids, aiming to reduce the training time without hurting the achieved reconstruction quality. 

Tab.~\ref{tab:different_grid_sizes} shows one set of our validation experiments in terms of the achieved reconstruction PSNR vs. the measured training time, where we vary the density grid size $S_D$ and color grid size $S_C$ from $S_D:S_C=0.25:1$ to $S_D:S_C=1:0.25$. Here we make two observations. 
(1) Adopting \textbf{different grid sizes for the color and density grids} leads to a better PSNR vs. training runtime trade-off, as compared to adopting the same grid size for both grids as in Instant-NGP~\cite{muller2022instant}. In particular, doing so reduces the training runtime by 12.5\% while maintaining the same PSNR as compared to Instant-NGP~\cite{muller2022instant}. 
(2) \textbf{Color features are less sensitive than density features} when their grid size is reduced. Particularly, when the color grid size or density grid size is reduced to 0.25$\times$ of the vanilla one in Instant-NGP~\cite{muller2022instant}, the achieved reconstruction PSNR is 26.0 dB vs. 25.4 dB, respectively, indicating the lower sensitivity of the former to the grid size (i.e., a higher level of \textbf{spatial redundancy}).

\begin{figure}[b]
\vspace{-2em}
  \centering
  \includegraphics[width=1\linewidth]{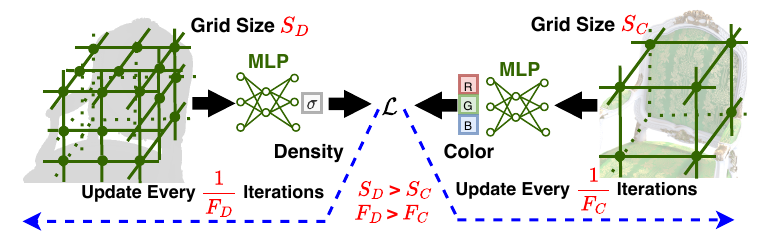}
\vspace{-2.5em}
\caption{Overview of the proposed Instant-3D algorithm pipeline with the decomposed color and density branches.}
\label{fig:alg_overview}
\vspace{-0.7em}
\end{figure}

\textbf{Proposed Technique.} Built upon our discovery above and consistent observation, we propose to use a smaller grid size for the color grid than that for the density grid, i.e., we ensure $S_D>S_C$ in our {\FrameworkName} algorithm as visualized in Fig.~\ref{fig:alg_overview}.

\vspace{-0.5em}
\subsection{Different Update Frequencies for the Color and Density Branches}
\label{sec:different_update_frequencies}

\begin{wraptable}{r}{0.5\linewidth}
\vspace{-1.5em}
\caption{The achieved reconstruction quality (PSNR) vs. the required training time on the eight scenes of NeRF-Synthetic~\cite{mildenhall2020nerf}, when adopting \textbf{different update frequencies} for the density grid $F_D$ and the color grid $F_C$. Here the training runtime is measured on an edge GPU~\cite{xavier_nx}.}
\vspace{-0.5em}
\centering
  \resizebox{1.0\linewidth}{!}
  {
    \begin{tabular}{c||c||c}
    \toprule
    \multirow{2}{*}{$F_D:F_C$} & Average Training & Average \\
     & Runtime (s) & Test PSNR \\
    \midrule
    1:1~\cite{muller2022instant} & 72 & 26.0 \\
    \midrule
    0.5:1  & 67 ($\downarrow$ 6.9\%) & 24.3 \\
    \textbf{1:0.5} & \textbf{65} ($\mathbf{\downarrow}$ \textbf{9.7\%}) & \textbf{25.9} \\
    \bottomrule
    \end{tabular}
    }
  \label{tab:different_update_frequencies}
\vspace{-1em}
\end{wraptable}

\textbf{Observations.} With similar motivation, we propose to adopt different update frequencies for the resulting color and density branches of the decomposed embedding grids to reduce the training time without hurting the reconstruction quality. 

Tab.~\ref{tab:different_update_frequencies} shows one set of our validation experiments in terms of the achieved reconstruction PSNR vs. the measured training time, where we vary the update frequencies for the density grid $F_D$ and color grid $F_C$ from $F_D:F_C=0.5:1$ to $F_D:F_C=1:0.5$. Similarly,  we can make two observations. (1) Adopting \textbf{different update frequencies for the color and density} grids leads to a better PSNR vs. training runtime trade-off, as compared to adopting the same update frequency for both as in Instant-NGP~\cite{muller2022instant}. Specifically, doing so reduces the training runtime by 9.7\% while keeping a similar PSNR as compared to Instant-NGP~\cite{muller2022instant}. 
(2) \textbf{Color features are less sensitive than density features} when the update frequency is reduced. In particular, 
when the update frequency for the color or density grid is reduced to 0.5$\times$ of the vanilla one in Instant-NGP~\cite{muller2022instant}, the achieved reconstruction PSNR is 25.9 dB vs. 24.3 dB, respectively, indicating the lower sensitivity of the former to the update frequency (i.e., a higher level of \textbf{temporal redundancy}). 

\textbf{Proposed Technique.} Similarly, we propose to use a lower update frequency for the color grid than that for the density grid, i.e., $F_D>F_C$ in our {\FrameworkName} algorithm (see Fig.~\ref{fig:alg_overview}).

\begin{figure}[b]
  \centering
\vspace{-0.5em}
  \includegraphics[width=0.85\linewidth]{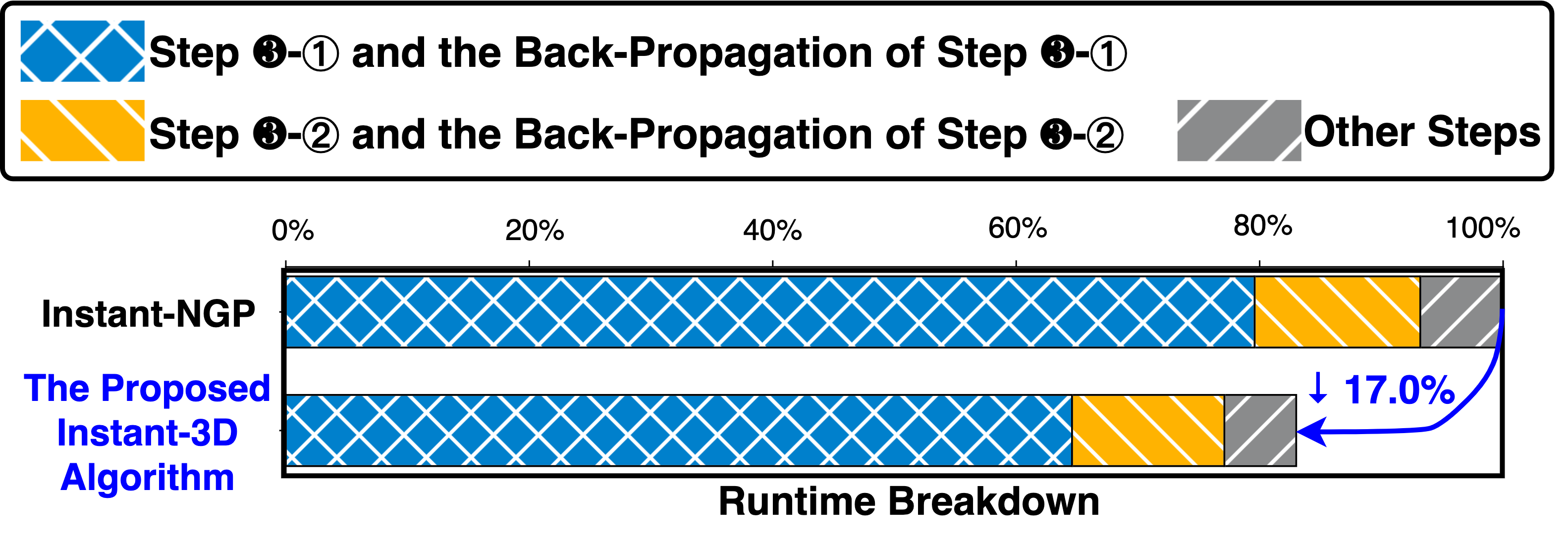}
\vspace{-1.5em}
\caption{Although the proposed Instant-3D algorithm can further accelerate the most efficient NeRF training algorithm~\cite{muller2022instant} by 17.0\% on average, the runtime breakdown averaged on the eight scenes of NeRF-Synthetic~\cite{mildenhall2020nerf} on edge GPU Xavier NX~\cite{xavier_nx} suggests that Step {\color{black}{\ding{184}}}-{\color{black}{\ding{172}}} (i.e., interpolating embeddings from the embedding grid) still dominates the training runtime of the proposed Instant-3D algorithm.}
\label{fig:profile_after_alg}
\end{figure}

\section{{\FrameworkName}: Proposed Accelerator}
\label{sec:hardware}

In this section, we first profile Instant-NGP~\cite{muller2022instant} when applying our proposed algorithm (see Sec. \ref{sec:acce-motivation}), showing that the achieved training time is still not satisfactory for instant on-device NeRF training. After that, we further analyze the memory access patterns of the dominant embedding grid interpolation during both the feed-forward and back-propagation processes when training Instant-NGP~\cite{muller2022instant} using our algorithm in Sec.~\ref{sec:HP_Analysis}. Finally, we present our proposed {\FrameworkName} accelerator in Sec.~\ref{sec:HW_design}, where the featured components include (1) a feed-forward read mapper to make good use of the on-chip multi-bank SRAM arrays (see Sec.~\ref{sec:HW_intra}), (2) a back-propagation update merger to minimize the number of required SRAM writes (see Sec.~\ref{sec:HW_inter}), and (3) a multi-core-fusion-based reconfigurable scheme to support the different grid sizes needed by our Instant-3D algorithm (see Sec.~\ref{sec:HW_reconf}).

\subsection{Motivation: Profiling Instant-NGP with Our Algorithm}
\label{sec:acce-motivation}

Fig.~\ref{fig:profile_after_alg} depicts the profiling results of our proposed algorithm, described in Sec.~\ref{sec:alg}, which accelerates the most efficient NeRF training algorithm Instant-NGP~\cite{muller2022instant} by 17.0\% on average.
However, the required training time per scene to achieve the satisfactory average PSNR of $>$ 25 dB~\cite{thomos2005optimized,li2007robust} is still around 60 seconds when being executed on edge GPU Xavier NX~\cite{xavier_nx}, as shown in Tab.~\ref{tab:different_grid_sizes} and Tab.~\ref{tab:different_update_frequencies}. Thus, the achievable training runtime is still far from the desired instant (i.e., $<$ 5 seconds~\cite{muller2022instant}) on-device 3D reconstruction. 
From Fig.~\ref{fig:profile_after_alg}, we observe that Step {\color{black}{\ding{184}}}-{\color{black}{\ding{172}}} (interpolating embeddings from the embedding grid) and its corresponding back-propagation process still dominate the training runtime, accounting for around 80\% of the total training runtime, motivating us to develop a dedicated accelerator to further boost the training efficiency for achieving instant on-device NeRF-based reconstruction.

\subsection{Analyzing the Memory Access Patterns During Training}
\label{sec:HP_Analysis}

\begin{figure}[b]
  \centering
  \vspace{-1em}
\includegraphics[width=1\linewidth]{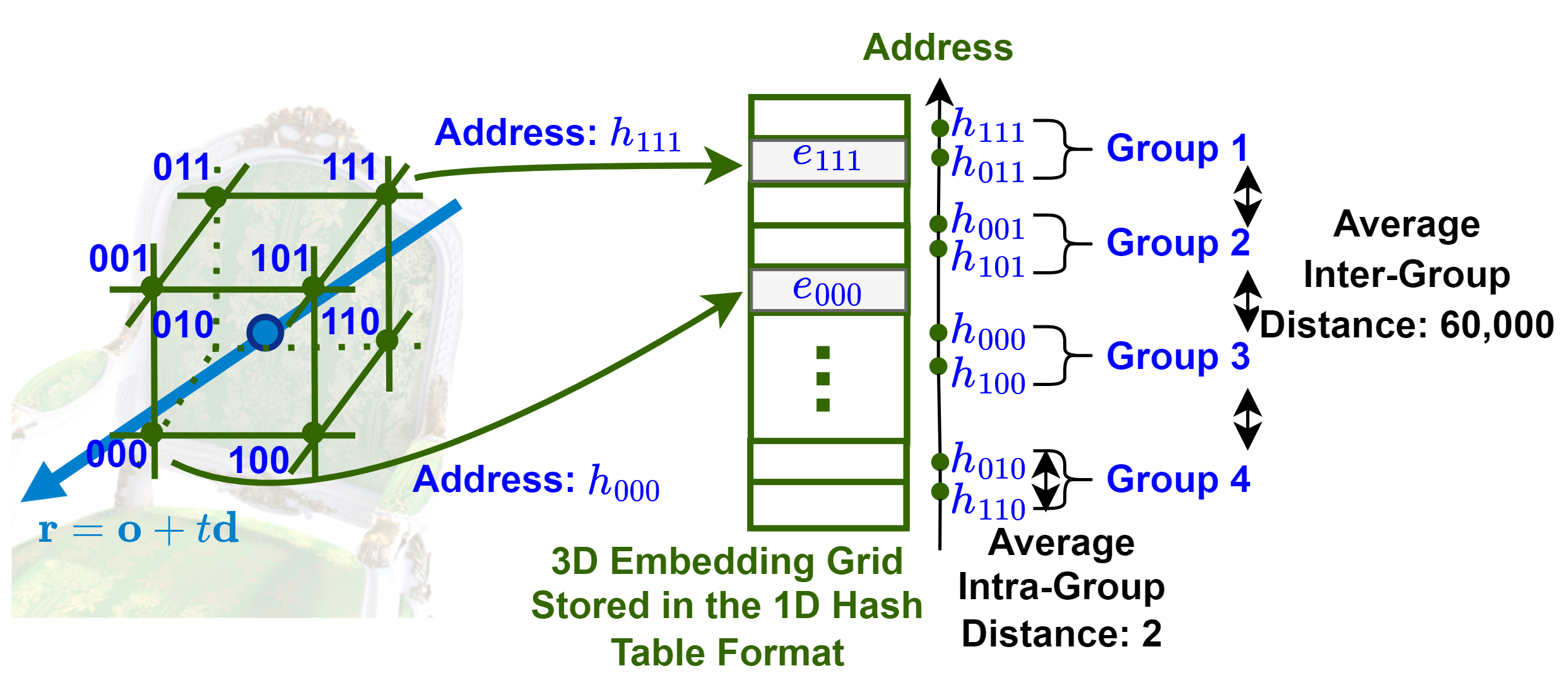}
\vspace{-2em}
\caption{The addresses for the embeddings of the eight neighboring vertices during embedding grid interpolation can be clustered into four groups, which is consistently observed on the eight scenes of NeRF-Synthetic~\cite{mildenhall2020nerf}.}
\label{fig:address_group}
\end{figure}

\begin{figure}[t]
  \centering  \includegraphics[width=1.0\linewidth]{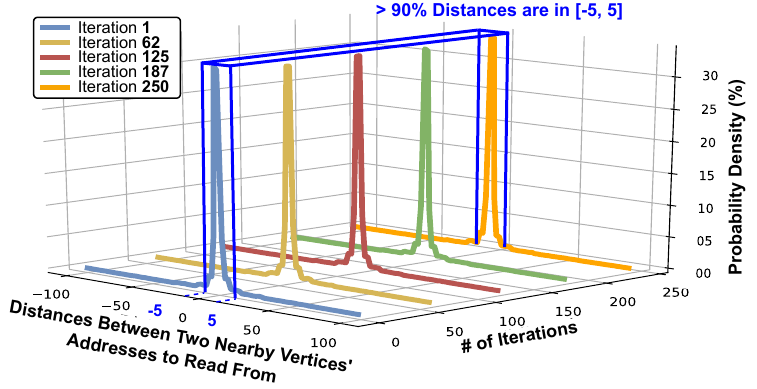}
\vspace{-1.5em}
\caption{More than 90\% of distances between the accessed addresses of neighboring vertices are between [-5, 5] when observing the memory access patterns during the training of the eight scenes in NeRF-Synthetic~\cite{mildenhall2020nerf}.}
\label{fig:hw_uc_IntraCount}
\vspace{-1.em}
\end{figure}

\textbf{Memory Access Patterns During Feed-Forward.}
\label{sec:HP_Analysis_intra}
Motivated by the observation that memory capacity is the bottleneck of runtime in~\cite{muller2022instant}, we further analyze the memory access patterns during the embedding grid interpolation that dominates the total training runtime. As illustrated in Sec.~\ref{sec:nerf_preliminaries}, to obtain the embeddings of each queried point along the camera rays passing through the pixels of the training images, the embeddings of the eight nearest surrounding vertices of the queried point are read from the 3D embedding grid.
Multiple 3D points can share the same cube of the 3D embedding grid; thus, their gradients will back-propagate to the same embedding of the cube, indicating the opportunity to merge those memory accesses to the same or similar addresses.
Specifically, we inspect all accessed addresses of the embedding grid during the feed-forward process by clustering the eight surrounding vertices into four groups, each of which contains two vertices with the same y-axis and z-axis, as illustrated in Fig.~\ref{fig:address_group}. Through extensive measurements on eight scenes~\cite{mildenhall2020nerf}, we find that (1) the inter-group distances can be very large (the average distance is as high as 60,000); and (2) about 90\% of intra-group distances are $<$ 5, as shown in Fig.~\ref{fig:hw_uc_IntraCount}. Both phenomena are consistently observed across different training iterations.

We conjecture that the above two memory access patterns are caused by the \textbf{remoteness} and \textbf{locality} of memory accessing, respectively.
Specifically, after passing the coordinates of the surrounding vertices as the inputs to the hash function in Eq.~\ref{eq:hash_func}, the output addresses of the embedding grid can be classified into the following two cases. \ul{Case 1:} If the differences between the coordinates of the surrounding vertices happen on the y-axis or z-axis, then such differences will be amplified by $\pi_2 = 2654435761$ or $\pi_3 = 805459861$, respectively, according to Eq.~\ref{eq:hash_func}. Such \textbf{remoteness} results in quite different addresses among the different groups as observed above (e.g., 60,000 on average, as shown in Fig.~\ref{fig:address_group}). \ul{Case 2:} If the aforementioned differences of coordinates happen on the x-axis, then such differences will not be amplified because of $\pi_1 = 1$ in Eq.~\ref{eq:hash_func}. Such \textbf{locality} results in similar addresses in the same group as observed above (e.g., 90\% of the address distances are $<$ 5, as shown in Fig.~\ref{fig:hw_uc_IntraCount}).

Motivated by (1) the fact that the 1D hash table, which represents the 3D embedding grid, is stored in multi-bank SRAM arrays due to the limited SRAM cell's size and (2) the aforementioned unique memory access patterns of the embedding grid during the feed-forward process, we propose a feed-forward read mapper to map the memory read requests of multiple nearby points' embeddings without bank access collisions into one read request, aiming to improve the resource utilization of the multi-bank SRAM arrays. We introduce the proposed detailed design of the feed-forward read mapper in Sec.~\ref{sec:HW_intra}.

\newcounter{TempEqCnt}
\setcounter{TempEqCnt}{\value{figure}} 
\setcounter{figure}{9} 
\begin{figure}[b]
  \centering
  \vspace{-1em}
  \includegraphics[width=1.0\linewidth]{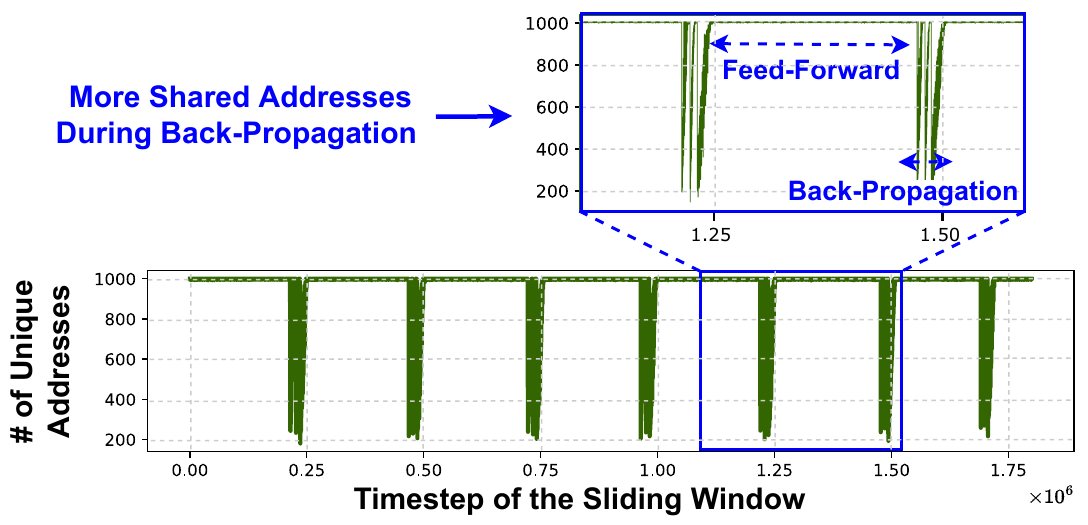}
\vspace{-2em}
\caption{The number of unique accessed addresses within a sliding window of 1000 continuous accesses indicates that there are fewer unique accessed addresses during the back-propagation process, enabling the opportunity to merge the accesses to those addresses of the shared embeddings.
}
\label{fig:hw_uc_InterUniqueness}
\end{figure}
\setcounter{figure}{\value{TempEqCnt}}

\setcounter{figure}{10}
\begin{figure*}[!t]
  \centering
  \includegraphics[width=1\linewidth]{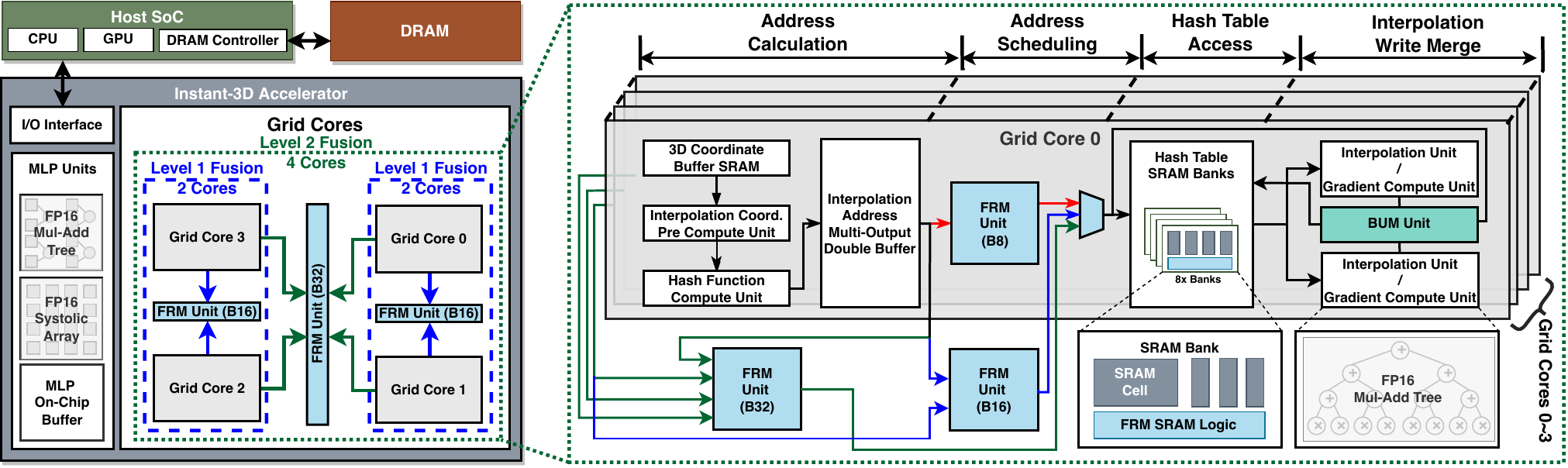}
\vspace{-1.5em}
\caption{Overall architecture of our proposed Instant-3D accelerator. There are 3 different modes corresponding to different hash table sizes: \ul{(1)} For a hash table size of 256 KB, \textcolor{red}{\textbf{Level 0 standalone mode (marked as red)}} is activated. In this mode, the four grid cores run independently, and the SRAM access of each core is managed by its internal FRM unit (B8); \ul{(2)} For a hash table size of 512 KB, \textcolor{blue}{\textbf{Level 1 Fusion mode (marked as blue)}} is activated, during which two grid cores are fused by scheduling the SRAM access of the two cores in the shared FRM Unit (B16); \ul{(3)} For a hash table size of 1MB, \textcolor{my_green}{\textbf{Level 2 Fusion mode (marked as green)}} is activated, during which all the 4 grid cores are fused together, and the SRAM access is managed by one FRM Unit (B32). Here the "B8/B16/B32" stands for 8/16/32 SRAM banks.}
\label{fig:hw_overall}
\vspace{-0.5em}
\end{figure*}

\textbf{Memory Access Patterns During Back-Propagation.}
\label{sec:HP_Analysis_inter}
As indicated by Eq.~\ref{eq:hash_func}, when the size of the 1D hash table that stores the 3D embedding grid is larger than the number of vertices in the grid, multiple vertices can share the same stored embeddings in the 1D hash table. This indicates an opportunity to reduce the memory accesses by merging multiple accesses of such shared embeddings into only one. To verify whether such shared embeddings are common in the bottleneck step of embedding grid interpolation, we analyze the number of unique accessed addresses within a small sliding time window (e.g., within 1000 continuous accesses) in Fig.~\ref{fig:hw_uc_InterUniqueness}. We can observe that (1) the number of unique accessed addresses varies along the training process and features predictable access patterns during the feed-forward and back-propagation processes; (2) for the access patterns during the feed-forward process, the aforementioned cases of shared embeddings do not exist, i.e., all of the 1000 continuous accesses in the sliding window are unique; (3) for the access patterns during the back-propagation process, there exists the cases of shared embeddings among more than five accesses from different timesteps, where $\sim$200 unique accesses are observed among the 1000 continuous accesses. 

Based on such consistently observed patterns of shared embeddings during the back-propagation process, we propose a back-propagation update merger to merge the embedding grid updates that share the same embedding addresses but in different timesteps into one update operation. We present the detailed design of the proposed back-propagation update merger in Sec.~\ref{sec:HW_inter}.

\subsection{Proposed Instant-3D Accelerator}
\label{sec:HW_design}
\textbf{Overview of Our Proposed Instant-3D Accelerator.}
\label{sec:HW_overview}
Aiming to design a complete acceleration system for NeRF-based 3D reconstruction, our system includes Dynamic Random Access Memory (DRAM), a host System on Chip (SoC), and our proposed accelerator that consists of three major components: the I/O interface, the MLP units, and the grid cores, as shown in Fig.~\ref{fig:hw_overall}. Specifically, during the feed-forward process of each training iteration, the host SoC first performs Step {\color{black}{\ding{182}}} (i.e., randomly sampling pixels as a batch) and Step {\color{black}{\ding{183}}} (i.e., mapping pixels to rays) in the NeRF training pipeline demonstrated in Sec.~\ref{sec:nerf_preliminaries}. After that, in Step {\color{black}{\ding{184}}} (i.e., querying the features of points along the rays), the coordinates of those queried points along the rays that pass through the pixels of the training images are applied to our proposed Instant-3D accelerator. Inside our Instant-3D accelerator, Step {\color{black}{\ding{184}}}-{\color{black}{\ding{172}}} (i.e., interpolating embeddings from the embedding grid) and Step {\color{black}{\ding{184}}}-{\color{black}{\ding{173}}} (i.e., computing the features of the queried points) are accelerated by the grid cores and MLP units, respectively. Finally, the features (i.e., color and density) of the queried points outputted by our Instant-3D accelerator are fed back to the DRAM, and then the remaining Step {\color{black}{\ding{185}}} (i.e., predicting pixels' color) and Step {\color{black}{\ding{186}}} (i.e., computing the reconstruction loss) are performed by the host SoC. The corresponding back-propagation process follows the same workload assignment as the aforementioned feed-forward process, e.g., the back-propagation of Step {\color{black}{\ding{184}}}-{\color{black}{\ding{172}}} and Step {\color{black}{\ding{184}}}-{\color{black}{\ding{173}}}, which are identified as the bottleneck step of NeRF training in Sec.~\ref{sec:SW_ProfilingInstNGP}, are performed on the proposed Instant-3D accelerator by the grid cores and MLP units, respectively. We provide detailed descriptions of our grid core and MLP unit designs as follows.

\textbf{Grid Core Design.} To perform the embedding grid interpolation and its corresponding back-propagation process of Step {\color{black}{\ding{184}}}-{\color{black}{\ding{172}}}, our Instant-3D accelerator consists of four grid cores, four of our proposed Back-Propagation Update Merger (BUM) units (i.e., 1 BUM unit per grid core), and seven of our proposed Feed-Forward Read Mapper (FRM) units (i.e., 4 FRM unit inside grid cores and 3 FRM unit among grid cores). In particular,
to perform Step {\color{black}{\ding{184}}} (i.e., interpolating embeddings from the embedding grid) with the grid cores in our Instant-3D accelerator, the components in the grid cores adopt the following order to execute the feed-forward process of this step: (1) The 3D Coordinate Buffer SRAM first caches all the 3D coordinates of the queried points along the rays that pass through the pixels of the current batch's training images; (2) The coordinates of each queried points' nearest eight vertices in the grid are calculated by \textit{the Interpolation Coord. Pre Compute Unit}; (3) The aforementioned eight vertices' coordinates are applied to \textit{the Hash Function Compute Unit} to execute the hash function (Eq.~\ref{eq:hash_func}) and output the corresponding eight addresses of the embedding grid stored in a 1D hash table format; (4) The addresses of the embedding grid to be accessed are fed into \textit{the Interpolation Address Multi-Output Double Buffer}; (5) The FRM units (see Sec.~\ref{sec:HW_intra}) first map multiple embedding grid read requests into fewer ones without causing memory bank access collisions and then send those requests to the Hash Table SRAM Banks to fetch the corresponding embeddings; (6) Finally, the fetched embeddings are applied into \textit{the Interpolation Unit} to obtain the trilinear interpolated values for generating the embeddings of the queried points. During the back-propagation of Step {\color{black}{\ding{184}}}, all the aforementioned components, except the Interpolation Unit, perform the same workload as the aforementioned feed-forward process.
Meanwhile, the Interpolation Unit is reconfigured into the Gradient Computation Unit for the calculation of the gradients of each accessed embeddings in the current batch. Moreover, to reduce memory accesses during the aforementioned action of writing back the updated embeddings, we propose to include the BUM unit (see Sec.~\ref{sec:HW_inter}) in each grid core to merge the updates of the same address into one update operation.

\textbf{MLP Unit Design.} To perform the feed-forward and back propagation process of MLP in Step {\color{black}{\ding{184}}}-{\color{black}{\ding{173}}}, we adopt two types of computing unit for them: (1) a systolic array MLP Unit and (2) a multiplier-adder-tree MLP Unit, dedicated to matrix multiplications with (1) a relatively large output channel (e.g., $>$ 3) and (2) a relatively small output channel (e.g., $\leq$ 3), respectively. Such a design with two unit types is inspired by the observations in~\cite{9793397,rao2022icarus}, showing that the multiplier-adder-tree can achieve a higher hardware utilization than the systolic array under the cases with relatively small output channels (e.g., $\leq$ 3). 

\begin{figure}[!b]
\vspace{-1.2em}
  \centering
  \includegraphics[width=0.9\linewidth]{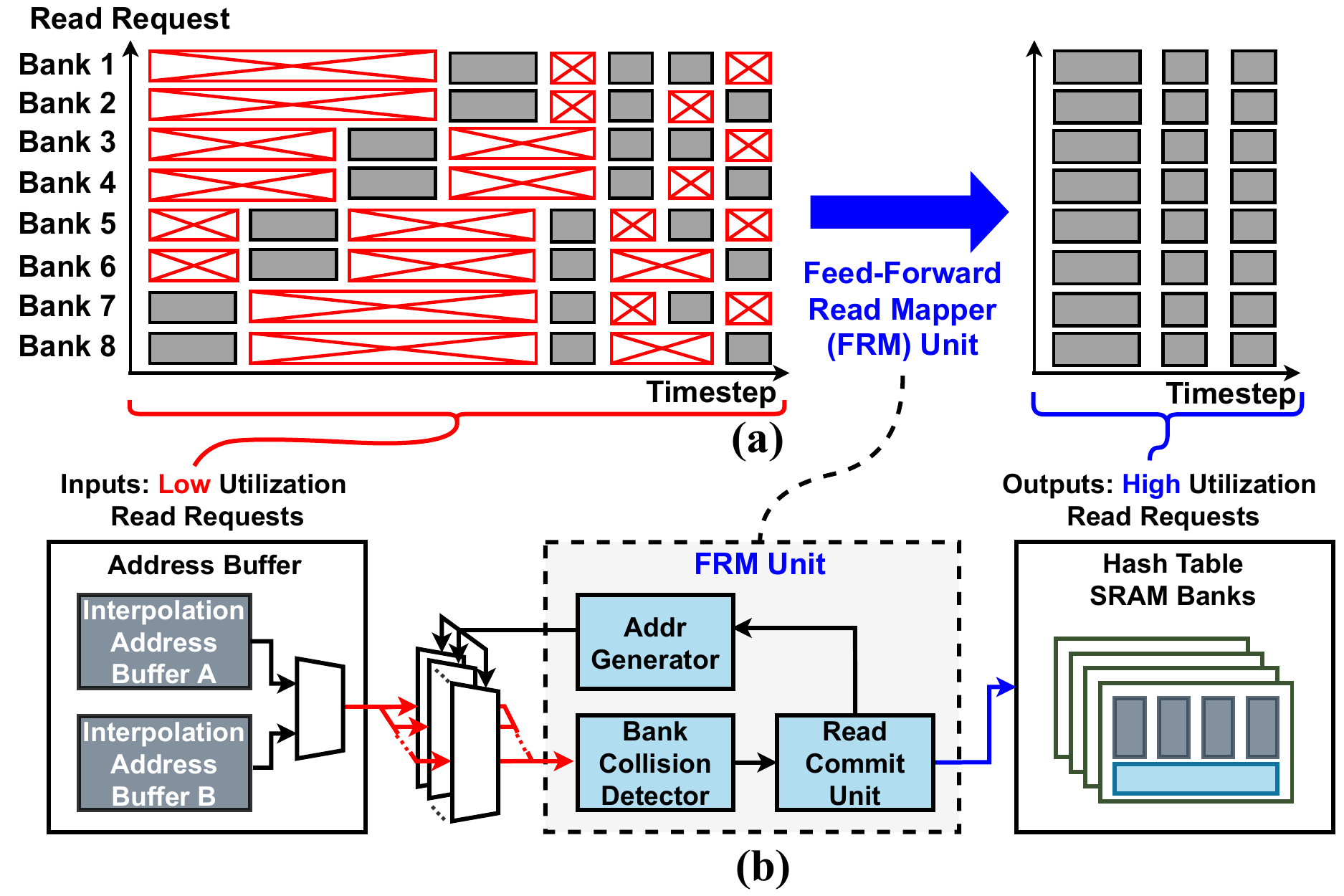}
\vspace{-1.7em}
\caption{The proposed FRM unit maps low utilization read requests to high utilization ones, as visualized in its (a) dataflow and implemented by its (b) hardware schematic.}
\vspace{-0.5em}
\label{fig:Read}
\end{figure}

\begin{figure*}[!b]
\vspace{-0.5em}
  \centering
\includegraphics[width=1.0\linewidth]{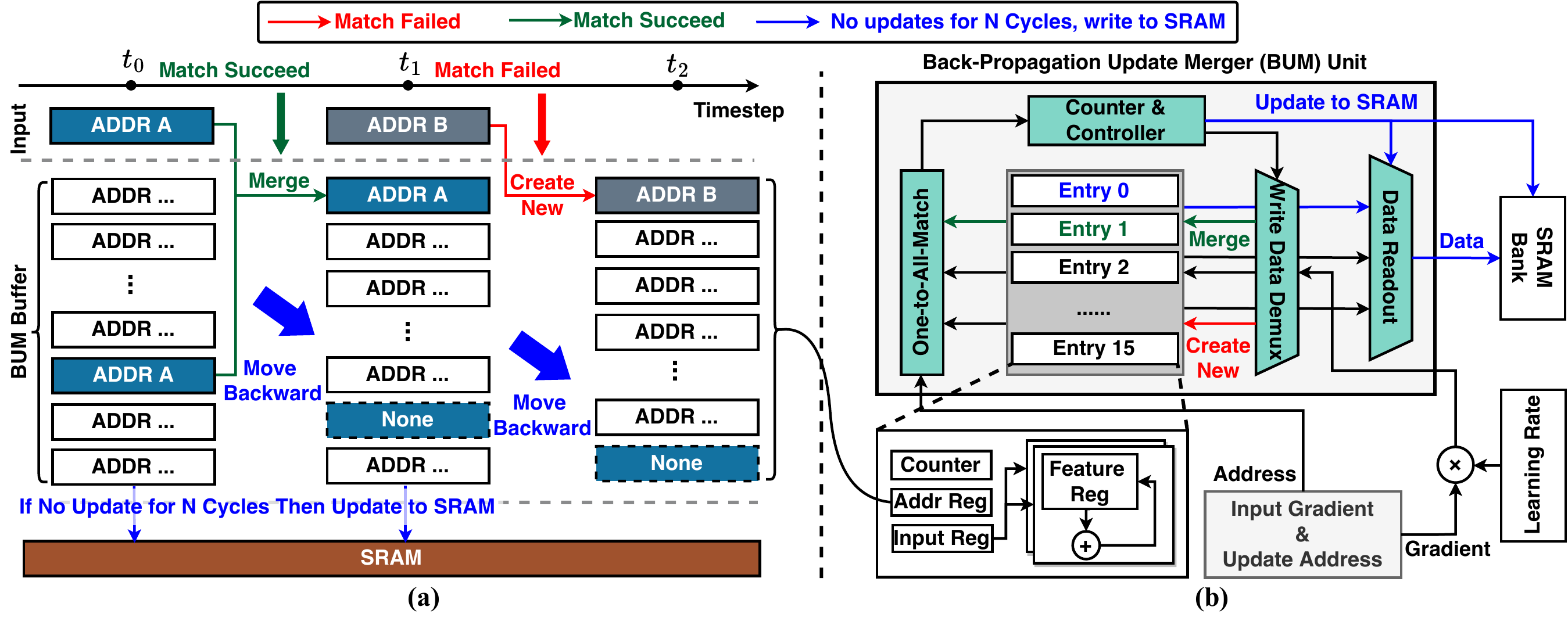}
\vspace{-2.5em}
\caption{The proposed BUM unit merges the update operations to the same embeddings but at different timesteps into one update operation, as visualized in its (a) dataflow and implemented by its (b) hardware schematic. \textcolor{red}{\textbf{Case 1 (marked as red):}} A match fails between the input address and any cached address in the BUM buffer (e.g., Address B at timestep $t_1$), then BUM creates a new entry to store the input address and its corresponding values to be updated; \textcolor{my_green}{\textbf{Case 2 (marked as green)}}: A successful match (e.g., Address A at timestep $t_0$), then the corresponding update operations of the matched pair are merged into one. If any address in the BUM buffer has not been matched for $N$ cycles, its corresponding values to be updated will be written back to SRAM \textcolor{blue}{\textbf{(marked as blue)}}.}
\label{fig:Write}
\end{figure*}

\subsection{Feed-Forward Read Mapper to Better Use SRAM Arrays}
\label{sec:HW_intra}

As mentioned in Sec.~\ref{sec:HP_Analysis_intra}, to read all the eight embeddings of each queried point's nearest vertices in the embedding grid, we divide the whole 1D hash table that stores the 3D embedding grid into eight banks equally and consider 2 cells per bank. Thus, the four clustered groups of eight embeddings are observed to be distributed in four or two memory banks, which causes low utilization of the multi-bank SRAM arrays (i.e., $4/8= 50\%$ or $2/8= 25\%$ SRAM bank utilization, assuming 8 banks in total for this case). To improve the memory utilization, we propose the FRM unit to better utilize SRAM bandwidth during the feed-forward process.
Specifically, as shown in Fig.~\ref{fig:Read}(a), multiple SRAM read requests from different clock cycles can be mapped into one clock cycle when there is no bank access collision. To achieve the goal of improving memory utilization, as shown in Fig.~\ref{fig:Read}(b), the FRM unit is designed to first fetch a batch of addresses from the address buffer, then detect the bank access collisions of those addresses. After that, the FRM unit maps the memory read requests without collisions into one and finally sends those mapped requests to the SRAM banks.

\subsection{Back-Propagation Update Merger Minimizing SRAM Writes}
\label{sec:HW_inter}

During the back-propagation process of each training iteration, there can be multiple (e.g., more than five) update operations to the same embeddings of the embedding grid, because of the cases where multiple vertices in the grid share the same embeddings stored in the 1D hash table, as analyzed in Sec.~\ref{sec:HP_Analysis_inter}. Inspired by the design strategy of trading higher-cost memory storage/access for lower-cost computation in~\cite{chen2016eyeriss,zhao2020smartexchange}, we propose a BUM unit to trade higher-cost memory write accesses for the lower-cost operations of merging the updates by accumulating the update values first and then writing back to the embedding grid. 

Specifically, as shown in Fig~\ref{fig:Write}(a), with the proposed BUM unit, if the current input address matches any cached ones in the BUM buffer (i.e., the case of $t_0$ in Fig~\ref{fig:Write}(a)), the corresponding update operations will be merged into one by accumulating the values to be updated; If the current input address does not match any cached ones (i.e., the case of $t_1$ in Fig~\ref{fig:Write}(a)), then the input address will be inserted into the BUM buffer. Meanwhile, if any address in the BUM buffer reaches the tail of the BUM buffer, this address will be popped out, and the corresponding accumulated values to be updated will be one write request to the SRAM. 

In the proposed BUM unit shown in Fig~\ref{fig:Write}(b), for each input address and its corresponding gradient, the gradient is multiplied by the pre-set learning rate, and the address is first fed into the One-to-All-Match module to verify whether it matches the addresses in different entries of the BUM buffer; After that, the address is sent to the matched entry to perform the accumulation of the values to be updated. However, if there is no matched entry for the input address, an empty entry is used to store the input address and its corresponding values to be updated (i.e., the gradient multiplied by the learning rate). Because the size of the BUM buffer is fixed, a controller and a counter to count the timesteps to the last update values accumulation for each entry are included in the BUM unit to read out the accumulated values to be updated in the entry and write them to the SRAM when the counter exceeds a pre-set threshold. Thus, the proposed BUM unit with the aforementioned design can merge the multiple memory write accesses to the same address into a single one by accumulating the update values in the BUM buffer.

\subsection{Reconfigurable Scheme Supporting Different Grid Sizes}
\label{sec:HW_reconf}

To leverage the properties of our proposed Instant-3D algorithm, i.e., adopting different grid sizes and update frequencies for the decomposed color and density grids (see Sec.~\ref{sec:alg}), it is desirable for our Instant-3D accelerator to be scalable to different grid sizes and update frequencies. Meanwhile, it is worth noting that our  Instant-3D accelerator is naturally scalable to different update frequencies by skipping one back-propagation process every $(\frac{1}{1-F})$ iteration, where $F$ can denote the update frequencies of the color or density grid. Thus, the key is to support different grid sizes. 

\begin{figure}[!t]
  \centering
  \includegraphics[width=0.95\linewidth]{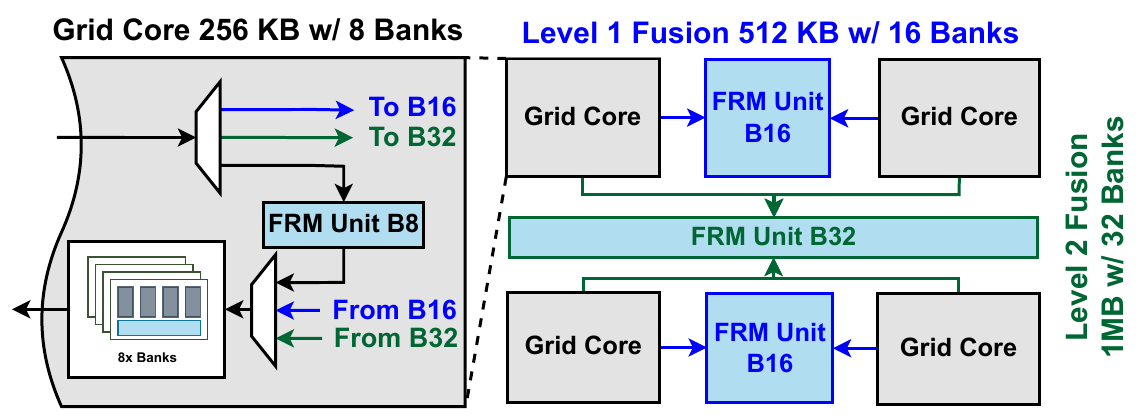}
\vspace{-0.5em}
\caption{A reconfigurable scheme for supporting different grid sizes in our Instant-3D algorithm in Sec.~\ref{sec:alg}, where B8/B16/B32 stands for 8/16/32 SRAM banks. Because different numbers of SRAM banks require different input sizes for enabling full utilization (e.g., 8/16/32 banks need 8/16/32 addresses), we design the FRM unit with different bank sizes for achieving full utilization of the SRAM bandwidth. }
\label{fig:ReconfHardware}
\vspace{-0.7em}
\end{figure}

To tackle the aforementioned challenge, we propose a multi-core-fusion-based reconfigurable scheme. As shown in Fig.~\ref{fig:ReconfHardware}, such a scheme is implemented by including a 16-bank FRM unit for each pair of grid cores to enable a Level 1 Fusion Mode for the pair (i.e., utilizing two grid cores in total to support a grid size of 512KB) and a 32-bank FRM unit between two pairs of the grid cores to make up a Level 2 Fusion Mode for the two pairs (i.e., utilizing four grid cores in total to support a grid size of 1MB).

\section{Evaluation}

In this section, we first introduce the detailed settings for evaluating our proposed Instant-3D framework in Sec.~\ref{sec:exp_settings}, and then benchmark our proposed Instant-3D algorithm and accelerator in Sec.~\ref{sec:exp_alg} and Sec.~\ref{sec:exp_acc}, respectively.

\begin{table}[b]
\vspace{-0.5em}
\caption{A summary of the considered devices' specifications.}
\centering
\vspace{-0.5em}
\setlength{\tabcolsep}{2pt}
  \resizebox{1.0\linewidth}{!}
  {
    \begin{tabular}{c||cccc}
    \toprule
    \textbf{Device} &  \textbf{Jetson Nano~\cite{jetson_nano}} & \textbf{Jetson TX2~\cite{tx2}} & \textbf{Xavier NX~\cite{xavier_nx}} &   \textbf{Instant-3D}  \\
    \midrule
    Technology & 20 nm & 16 nm & 12 nm & 28 nm \\ 
    
    \midrule
    SRAM  & 2.5 MB & 5 MB & 11 MB & 1.5 MB\\
    
    \midrule
    Area & 118 mm$^2$ & N/A & 350 mm$^2$ & 6.8 mm$^2$ \\
    
    \midrule
    Frequency & 0.9 GHz & 1.4 GHz & 1.1 GHz &  0.8 GHz \\
    
    \midrule
    DRAM & LPDDR4-1600 &  LPDDR4-1866 & LPDDR4-1866 &  LPDDR4-1866 \\
    Bandwidth & 25.6 GB/s &  59.7 GB/s & 59.7 GB/s &  59.7 GB/s \\
    
    \midrule
    Typical Power & 10 W & 15 W & 20 W  & 1.9 W\\
    
    \bottomrule
    \end{tabular}
    }
  \label{tab:bencmark_setting}
\end{table}

\subsection{Evaluation Settings}
\label{sec:exp_settings}

\textbf{Datasets \& Baselines.}
\underline{\textit{Datasets:}}
To evaluate the achieved reconstruction quality and training efficiency of our proposed Instant-3D, we conduct experiments on the commonly-used NeRF-Synthetic~\cite{mildenhall2020nerf}, the large-scale SILVR~\cite{courteaux2022silvr}, and the real-world-captured ScanNet~\cite{dai2017scannet} datasets.
The reconstruction quality is measured by the PSNR of the corresponding test set. 
\underline{\textit{Baselines:}}
We consider three commercial hardware devices as our baselines, including a Jetson Nano~\cite{jetson_nano} with a typical power consumption of 10 W, a Jetson TX2~\cite{tx2} with a typical power consumption of 15 W, and a Xavier NX~\cite{xavier_nx} with a typical power consumption of 20 W. 
In our experiments, the energy of the aforementioned baseline devices is measured using embedded power-rail monitors following~\cite{li2021hw}.
We emphasize that to the best of our knowledge, our proposed Instant-3D is the first to develop accelerators for NeRF-based 3D reconstruction training, and thus there are no dedicated NeRF training accelerator baselines for comparison.
The hardware specifications of all baselines and our Instant-3D are summarized in Tab.~\ref{tab:bencmark_setting}.
Note that we do not benchmark with RT-NeRF~\cite{li2022rt} and ICRUAS~\cite{rao2022icarus}, which are the prior works on NeRF acceleration, as they can only perform NeRF inference instead of NeRF training, failing to support the desired instant on-device 3D reconstruction.

\textbf{Instant-3D Algorithm Implementation.} The implementation of the proposed Instant-3D algorithm is based on the open-sourced Instant-NGP~\cite{muller2022instant}'s official CUDA implementation. 
We follow the default algorithm settings in Instant-NGP~\cite{muller2022instant} excepting for the density grid's size and update frequency, i.e., $S_D$ and $F_D$, and the color grid's size and update frequency, i.e., $S_C$ and $F_C$, which are set as $S_D:S_C=1:0.25$ and $F_D:F_C=1:0.5$.
Such a configuration is selected as the one that compresses the training cost most but also maintains the same reconstruction quality with Instant-NGP~\cite{muller2022instant} by performing a grid search from $1:0.125$, $1:0.25$, $1:0.5$, and $1:0.75$. 
Thus, to implement $S_D:S_C=1:0.25$, the 1D hash tables store the density grid and color grid that have $2^{16}$ and $2^{18}$ entries, respectively, and both have 2 features per entry, following Instant-NGP~\cite{muller2022instant}. 
Additionally, to implement $F_D:F_C=1:0.5$, the density grid is updated by the back-propagation of the reconstruction loss every iteration, and the color grid is updated every two iterations.

\begin{figure}[!b]
\vspace{-0.7em}
  \centering
\includegraphics[width=0.92\linewidth]{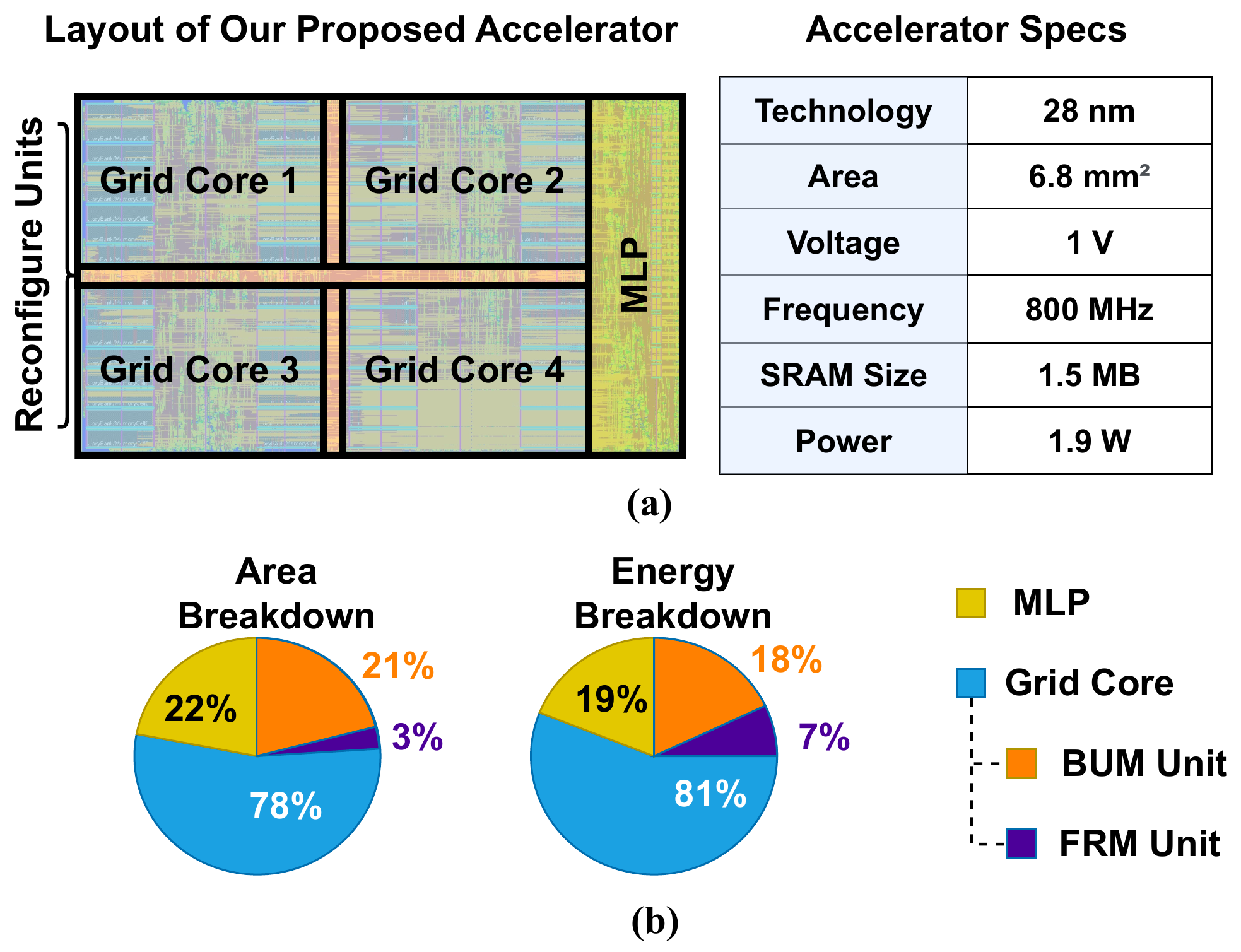}
\vspace{-1em}
\caption{(a) The layout and performance specifications and (b) the layout, energy, and area breakdown of our Instant-3D accelerator.}
\label{fig:layout}
\end{figure}

\begin{figure*}[!t]
  \centering
\includegraphics[width=1.0\linewidth]{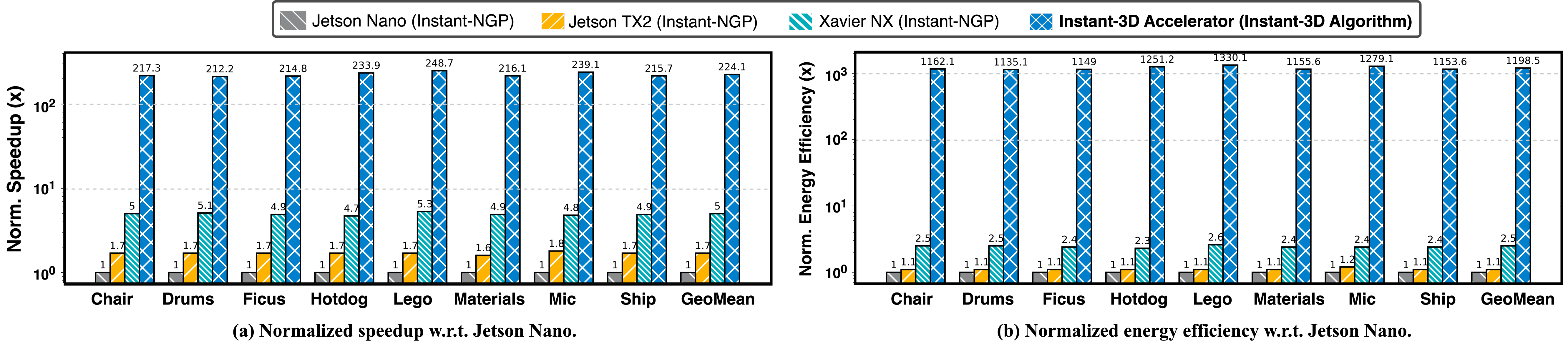}
\vspace{-2em}
\caption{The normalized speedup and energy efficiency achieved by our proposed Instant-3D and three baseline devices on the eight scenes of NeRF-Synthetic~\cite{mildenhall2020nerf}. The legends follow the ``device (algorithm)'' format.}
\vspace{-0.3em}
\label{fig:HW_Bench}
\end{figure*}

\textbf{Instant-3D Accelerator Implementation.} 
To evaluate the energy and the area of the proposed Instant-3D accelerator, we implement our accelerator in RTL, synthesize the RTL design using Synopsys Design Compiler~\cite{synopsys_design_compiler}, and then place \& route the design using Cadence Innovus~\cite{cadence_innovus}, based on a commercial 28nm CMOS technology.
In all our experiments, we set the reordering pipeline depth of our proposed FRM and BUM units to be 16, based on empirical observations and find it to be generally applicable to all datasets in our experiments; and each SRAM array connected to FRM and BUM units can handle eight unique memory accesses.
Specifically, we use 16-bit half-precision floating-point arithmetic for all algorithm-related computations to ensure minimal rendering quality degradation due to quantization. In addition, we develop a cycle-accurate simulator to estimate the training efficiency of our proposed Instant-3D accelerator on different datasets with the assumption of a 59.7 GB/s DRAM bandwidth, which is the same as the typical DRAM bandwidth in LPDDR4-1866 used in the baseline hardware devices, Jetson TX2~\cite{tx2} and Xavier NX~\cite{xavier_nx}. The proposed Instant-3D accelerator consumes an area of 6.8 mm$^2$ and an average power consumption of 1.9 W, depicted in 
Fig.~\ref{fig:layout}

\subsection{Instant-3D Algorithm's Performance}
\label{sec:exp_alg}

\textbf{Benchmark with the SOTA Efficient NeRF Training Algorithm.} To evaluate the effectiveness of our Instant-3D algorithm in Sec.~\ref{sec:alg}, we benchmark it with the most efficient NeRF training algorithm, Instant-NGP~\cite{muller2022instant}, in terms of the achieved reconstruction PSNR and training runtime on edge GPU, Xavier NX~\cite{xavier_nx}. As shown in Tab.~\ref{tab:algs_benchmark}, we can observe that our proposed Instant-3D algorithm surpasses the most efficient NeRF training algorithm~\cite{muller2022instant} in terms of the reconstruction quality vs. training runtime trade-offs, e.g., 60 seconds vs. 72 seconds to achieve the same quality (26.0 PNSR averaged on the 8 scenes of NeRF-Synthetic~\cite{mildenhall2020nerf}).

\begin{table}[b]
\vspace{-1em}
\caption{Benchmark our proposed Instant-3D  algorithm with the most efficient NeRF training algorithm~\cite{muller2022instant}, in terms of the PSNR and training runtime on edge GPU Xavier NX~\cite{xavier_nx}.}
\vspace{-0.5em}
\centering
\setlength{\tabcolsep}{2pt}
  \resizebox{1.0\linewidth}{!}
  {
    \begin{tabular}{c||ccc||ccc}
    \toprule
    \multirow{2}{*}{Methods} & \multicolumn{3}{c}{Avg. Train. Runtime} & \multicolumn{3}{c}{PSNR} \\
     & NeRF-Synthetic~\cite{mildenhall2020nerf} & SILVR~\cite{courteaux2022silvr} & ScanNet~\cite{dai2017scannet} & NeRF-Synthetic~\cite{mildenhall2020nerf} & SILVR~\cite{courteaux2022silvr} & ScanNet~\cite{dai2017scannet}  \\
    \midrule
    Instant-NGP~\cite{muller2022instant} & 72 sec. & 135 sec.  & 84 sec. & 26.0 & 25.0 & 24.9 \\
    \textbf{Instant-3D} & \textbf{60 sec.} & \textbf{111 sec.}   &  \textbf{72 sec.} & \textbf{26.0} & \textbf{25.1} & \textbf{25.1} \\
    \bottomrule
    \end{tabular}
    }
  \label{tab:algs_benchmark}
\end{table}

\begin{figure}[t]
\vspace{0.5em}
  \centering
  \includegraphics[width=1.0\linewidth]{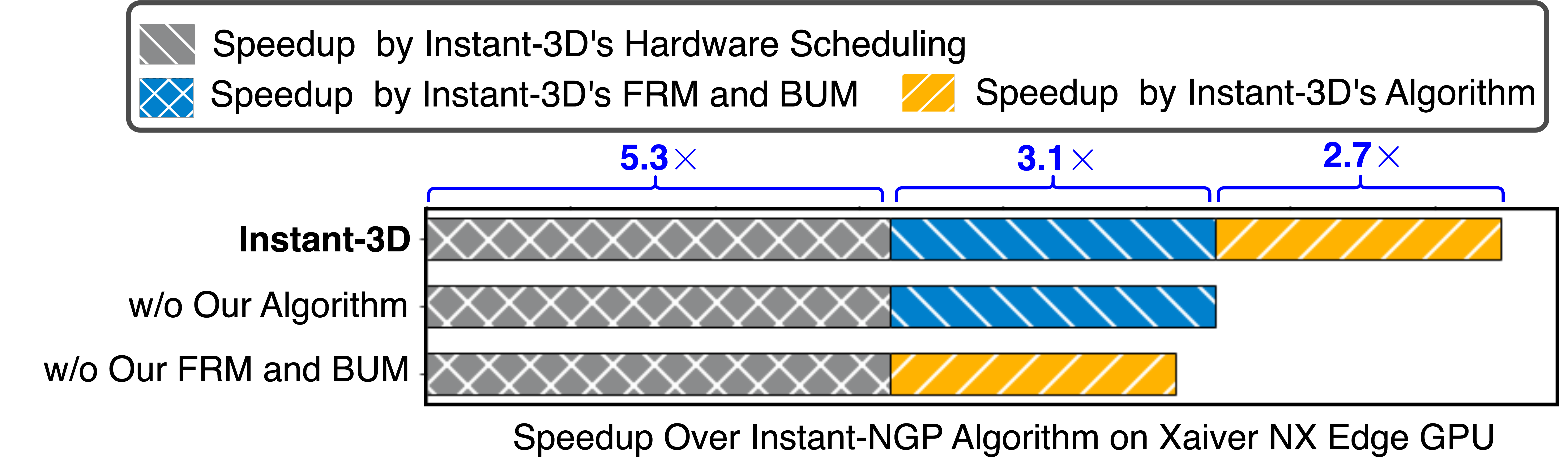}
\vspace{-2em}
\caption{The speedup (in logarithmic scale) over Instant-NGP~\cite{muller2022instant} algorithm on Xaiver NX Edge GPU~\cite{xavier_nx} achieved by different techniques of our proposed Instant-3D on NeRF-Synthetic dataset~\cite{mildenhall2020nerf}.}
\label{fig:speedup_breakdown}
\vspace{-1em}
\end{figure}

\subsection{Instant-3D Accelerator's Performance}
\label{sec:exp_acc}

\textbf{Benchmark with the SOTA Efficient NeRF Training Devices.}
We summarize the NeRF training efficiency improvements achieved by our proposed Instant-3D accelerator in Fig.~\ref{fig:HW_Bench}. Specifically, as compared to the three baselines on NeRF-Sythetic~\cite{mildenhall2020nerf}, the proposed Instant-3D accelerator offers on average 224$\times$/132$\times$/45$\times$ speedups and 1198$\times$/1089$\times$/479$\times$ more energy efficiency over Jetson Nano~\cite{jetson_nano}/Jetson TX2~\cite{tx2}/Xavier NX~\cite{xavier_nx}, respectively. 
Specifically, our Instant-3D’s 45$\times$ speedup over Xaiver NX~\cite{xavier_nx} results from (1) 2.7$\times$ speedup by the Instant-3D algorithm, (2) 3.1$\times$ speedup by our FRM and BUM units, which are inspired by the observed memory access patterns in Sec.~\ref{sec:HP_Analysis}, and (3) 5.3$\times$ speedup by better hardware scheduling, i.e., our multi-core-fusion-based-reconfigurable-scheme, as shown in Fig.~\ref{fig:speedup_breakdown}.

\begin{figure}[b]
  \centering
  \vspace{-1em}
  \includegraphics[width=1.0\linewidth]{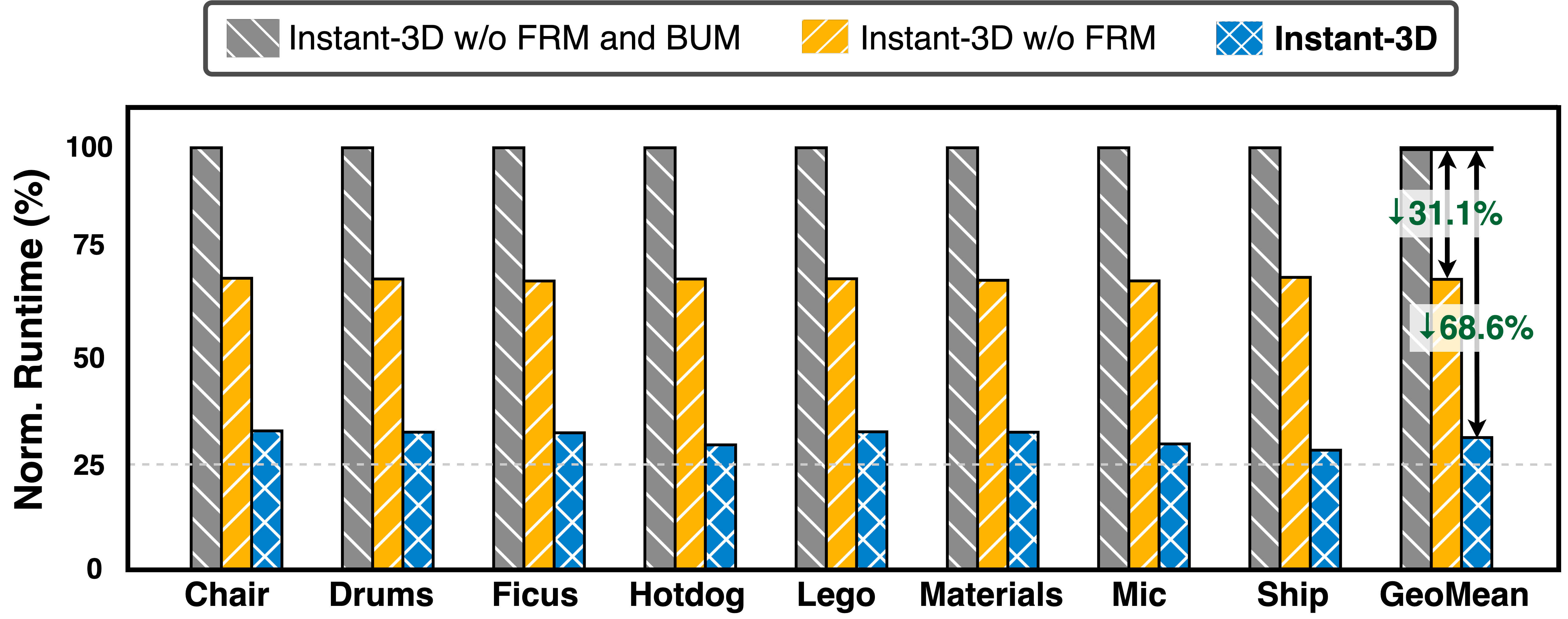}
\vspace{-2em}
\caption{The normalized runtime achieved by our proposed Instant-3D accelerator w/o the proposed FRM unit, depicted in Sec.~\ref{sec:HW_intra}, or w/o BUM unit, depicted in Sec.~\ref{sec:HW_inter}, on the eight scenes of NeRF-Synthetic~\cite{mildenhall2020nerf}.}
\label{fig:HW_Abl}
\end{figure}

\textbf{Ablation Study on the Effectiveness of the Proposed FRM and BUM Units.} As illustrated in the Sec.~\ref{sec:HW_intra} and Sec.~\ref{sec:HW_inter}, we propose an FRM unit to make good use of the on-chip multi-bank SRAM arrays and a BUM unit to minimize the number required SRAM writes. To verify the effectiveness of the two proposed units, we summarize the runtime of Instant-3D w/o the FRM unit or BUM unit in Fig.~\ref{fig:HW_Abl}. In particular, the proposed FRM unit can trim down the runtime by 31.1\% on average. Additionally, with the proposed BUM unit on top of the FRM unit, the runtime reduction ratio can be further enlarged to 68.6\%, which indicates that both the proposed FRM unit and BUM unit are necessary for our Instant-3D accelerator to achieve the desired instant on-device NeRF-based reconstruction. 
It is worth noting that, to the best of our knowledge, controlling the precise/fine-grained memory accesses required by the FRM and BUM units is not currently supported by CUDA's APIs on the benchmark devices summarized in Tab.~\ref{tab:bencmark_setting}.

\textbf{Ablation Study on Necessity of Co-Design.} To the best of our knowledge, our proposed Instant-3D, as an algorithm-hardware co-design acceleration framework, is the first that has achieved instant on-device NeRF-based 3D reconstruction. To verify the necessity of such a co-design strategy, we summarize the runtime of our Instant-3D w/o the proposed algorithm techniques or hardware techniques in Tab.~\ref{tab:exp_co_design}. Specifically, with our proposed Instant-3D algorithm, we can trim down the runtime by 83.0\% as compared to the most efficient NeRF training algorithm~\cite{muller2022instant} on the same edge GPU Xavier NX~\cite{xavier_nx}. Moreover, with both our proposed Instant-3D algorithm and accelerator, the runtime can be reduced to 2.2\% of the most efficient NeRF training solution (Instant-NGP~\cite{muller2022instant} on a Xavier NX~\cite{xavier_nx}), which indicates the necessity of the co-design strategy of our Instant-3D framework.

\begin{table}[!t]
\caption{The normalized runtime achieved by our Instant-3D framework w/o the proposed algorithm techniques or hardware techniques on different datasets.}
\centering
\vspace{-0.5em}
\setlength{\tabcolsep}{2pt}
  \resizebox{1\linewidth}{!}
  {
    \begin{tabular}{c||ccc}
    \toprule
    NeRF Training Solution  & \multicolumn{3}{c}{Normalized Runtime (\%) on} \\
    (Algorithm @ Hardware)  & NeRF-Synthetic~\cite{mildenhall2020nerf} & SILVR~\cite{courteaux2022silvr} & ScanNet~\cite{dai2017scannet} \\
    \midrule
    Instant-NGP~\cite{muller2022instant} @ Xavier NX~\cite{xavier_nx}& 100 & 100 & 100 \\
    \midrule
    \textbf{Instant-3D Algorithm} @ Xavier NX~\cite{xavier_nx}& 83.3 & 82.2 & 85.7 \\
    \midrule
    \textbf{Instant-3D Algorithm} @ \textbf{Instant-3D Accelerator} & 2.3 & 3.4 & 3.2 \\
    \bottomrule
    \end{tabular}
    }
  \label{tab:exp_co_design}
  \vspace{-1em}
\end{table}

\section{Related Works}

To the best of our knowledge, both of the only two existing works on designing dedicated accelerators for NeRF~\cite{rao2022icarus,li2022rt} can only perform NeRF inference, and thus cannot be adopted to achieve the goal of instant on-device 3D reconstruction.
  When compared with the SOTA NeRF inference accelerator~\cite{li2022rt}, our Instant-3D can achieve real-time ($>$ 30 FPS) rendering speed-up while only consuming 19.5\% of energy per frame and 36\% of the chip area. Moreover, Instant-3D achieves a 1,8002$\times$ speedup over an MLP-based NeRF inference accelerator~\cite{rao2022icarus}. It is worth noting that the prior works on MLP or Convolutional Neural Network (CNN) training acceleration~\cite{li2019tnpu,choi2020energy,qiao2018atomlayer} do not support the dominant operations of interpolating embeddings from the embedding grid, as analyzed in Sec.~\ref{sec:SW_ProfilingInstNGP}. Thus, they are not applicable to accelerate NeRF training.

\section{Conclusion}

We propose Instant-3D, which to the best of our knowledge is \textbf{the first} that has achieved instant on-device NeRF-based 3D reconstruction. Instant-3D algorithm decomposes the bottleneck embedding grid in terms of color and density to orthogonally squeeze out the redundancy in both branches; Instant-3D accelerator integrates an FRM unit to make good use of the on-chip multi-bank SRAM arrays, a BUM unit to minimize the number of required SRAM writes, and a reconfigurable scheme to support our instant-3D algorithm. We believe this work can open up an exciting perspective toward instant on-device 3D reconstruction for AR/VR.

\begin{acks}
This work was supported by the National Science Foundation (NSF) through two CCF programs (Award ID: 2211815 and 2312758) and CoCoSys, one of the seven centers in JUMP 2.0, a Semiconductor Research Corporation (SRC) program sponsored by DARPA.
\end{acks}

\bibliographystyle{ACM-Reference-Format}
\bibliography{sample-base}

\end{document}